# "Manifestation of Coupled Geometric Complexity in Urban Road Networks under Mono-Centric Assumption"

**Farideddin Peiravian[1,*] and Sybil Derrible[1,†]**

[1] University of Illinois at Chicago, Complex and Sustainable Urban Networks (CSUN) Lab, Civil Engineering Department, 842 W. Taylor St. (MC246), Chicago, IL, US 60607

[*] Research Associate, Email: fpeira2@uic.edu, Tel: 312-996-2429, Fax: 312-996-2426

[†] Assistant Professor, Email: derrible@uic.edu

**This article analyzes the complex geometry of urban transportation networks as a gateway to understanding their encompassing urban systems. Using a proposed ring-buffer approach and applying it to 50 urban areas in the United States, we measure road lengths in concentric rings from carefully-selected urban centers and study how the trends evolve as we move away from these centers. Overall, we find that the complexity of urban transportation networks is naturally coupled, consisting of two distinct patterns: (1) a fractal component (i.e., power law) that represent a uniform grid, and (2) a second component that can be exponential, power law, or logarithmic that captures changes in road density. From this second component, we introduce two new indices, density index and decay index, which jointly capture essential characteristics of urban systems and therefore can help us gain new insights into how cities evolve.**

Cities are complex systems consisting of many inter-related components and features. They possess both visible as well as hidden characteristics that are, similar to complex living organisms, lying beneath their physical forms. Moreover, the evolution and spread of an urban system and its components happen over many years, as the aggregated outcome of numerous individual and collective choices, each influenced by the prevailing conditions in its time. Each new change is overlaid on previous changes. In other words, any urban system and its components have a starting point when and where they are founded; tens or hundreds of years ago. While we may assume that the older a city is, the less coherent its founding blocks have been, many researchers suggest [1–3], and even demonstrate [4–8], that no matter how an urban system evolves or what foundations it is built on, from a larger perspective it has inherent order and organization.

As complex systems, cities have been studied heavily in the scientific literature, leading for a push towards a new "Science of Cities" [9]. In order to better understand the complex nature of an urban system, studies have focused on the characterization of its components [10]. As a result, the hidden, and presumably orderly, characteristics of different components of a given urban system have been a matter of interest in recent time [11–19]. These studies have looked at a large number of characteristics, from the travel behavior of their residents [20,21], to the amount of energy that is being consumed [22,23], to how they scale with size [2,24–28] to name a few.

From a methodology perspective, the complex behavior of cities has often been studied through their transportation systems [1,25] because "understanding the topology of urban networks that connect people and places leads to insights into how cities are organized" [25]. Urban transportation systems are particularly interesting to study since they have evolved at the same pace as their encompassing cities, and thus they offer virtual snapshots of the past through the changes in their characteristics from downtowns to the suburbs. Measuring the complex properties of transportation systems can therefore pave the way to a better understanding the formation and growth of cities.

In the case of an urban road network, one can visually observe that such an order manifests itself in a self-similar pattern [29,30]. In other words, the evolution of a transportation network is very similar to a tree that grows, then splits into branches, and those branches also grow and then split further into smaller branches, and so on so forth. One main difference, however, is that transportation networks create loops through branch-joining. Additionally, order can manifest itself by showing similar shapes and patterns even if scales differ. This is particularly true in road networks that tend to be denser in a downtown while keeping the same overall pattern throughout the city.

With the advent of new technologies, and in particular powerful Geographic Information System (GIS) tools, as well as the availability of more disaggregate datasets including extensive geospatial data, we are now able to perform a more detailed analysis of transportation networks towards a better understanding of their encompassing urban systems as complex adaptive entities.

Based on the above discussion, the main objective of this work is to employ a proposed ring-buffer approach to capture the complex geometric characteristics of urban road systems. The steps in the method section are first applied to Chicago metropolitan area as a leading example, and then applied to 50 U.S. metropolitan statistical areas (MSA). Overall, this work fits within the global endeavor to analyze cities and their infrastructure as complex systems [7,31,24,32–37]. Taking a complex analysis approach to better understand an urban system and its components offers many benefits, including the provision of measurable metrics, as is the case here.

**RESULTS**

Based on the proposed method, we decouple the total road length within a buffer of radius *r* around an urban center as the product of area and road density, as shown in equation 1:

$$N(r) = A \cdot \rho(r) = \pi r^2 \cdot \rho(r) \quad (1)$$

in which the road density, *ρ(r)*, can have any functional form. Despite its simple appearance, this relationship has a profound meaning. The first part of the right side, $\pi r^2$, is in fact a power law representing a fractal with the dimension of 2, i.e. a uniform grid. This means that a given road network does in fact have an intrinsic component similar to fractal features, which represents some kind of scaling in the road system and akin to many other studies [9,38]. Urban transportation systems, however, are not uniform, and road densities tend to decrease as we move away from the urban center. This property of road systems is captured in the second component, *ρ(r)*. What makes the above finding interesting is how these two components are coupled, which represents complexity at a higher level than what the power law alone offers. This presents the challenge of separating these two components, $\pi r^2$ and *ρ(r)*, thus isolating *ρ(r)*, in order to examine its form and characteristics.

The proposed ring-buffer method is applied to 50 U.S. urban road networks. A complete list of the urban areas studied, as well as the corresponding results, is provided in the supplementary materials. The results clearly show that there are three possible trends for the second component, representing different forms of the complexity of urban road networks: (1) exponential, (2) power law, and (3) logarithmic. Overall, the results confirm that the methodology developed in this study is sound, efficient, and robust.

For the exponential trend, as illustrated by the case of the Chicago Metropolitan Statistical Area (MSA) road network (Figure 1a), the total road length *N* within a buffer of radius *r* can be written in the following general form:

$$N(r) = \pi r^2 \cdot a \cdot e^{-br} \quad (2)$$

The second component in equation 2, i.e., $a \cdot e^{-br}$, has an exponential form and represents the road density of the network. We see that *a* is the maximum road density of the network (at *r* = 0). This also means that the larger *a* is, the larger *N* value will be, i.e. they have a direct and positive linear relationship. In other words, given *r* and *b* values, *a* represents how compact or sparse the road network is. Because of that, *a* can be considered as an overall *compactness* or *density index* for the urban road networks. In comparison, *b* has a different impact on *N*. We note that *b* appears with a negative sign in the exponent of the exponential function. This means that for a given set of *a* and *r* values, the larger *b* is, the smaller *N* will be. On the other hand, and unlike *a*, *b* has a non-linear inverse impact on *N*. As a result, *b* can be interpreted as a *decay index*. Put differently, *b* measures how fast the road density drops.

For the power law trend, e.g., the case of Austin MSA road network (Figure 1b), the total road length *N* within a buffer of radius *r* can be written in the following general form:

$$N(r) = \pi r^2 \cdot a \cdot r^{-b} \tag{3}$$

Similar to equations 1 and 2, the $\pi r^2$ represents a uniform grid, which is essentially the hidden fractal nature of road network under study. The second part, i.e., $a \cdot r^{-b}$, that represents the road density of the network, is also a form of power law. Similar to the exponential function discussed before, here *a* and *b* represent the density and decay indices, respectively.

For the logarithmic trend, e.g. the case of Los Angeles MSA road network (Figure 1c), the general form for the total road length *N* within a buffer of radius *r* can be written as:

$$N(r) = \pi r^2 \cdot [-b \cdot ln(r) + a] \tag{4}$$

Using the same analogy as equations 2 and 3, the $\pi r^2$ part represents a uniform grid, and *a* and *b* in the second component represent the density and decay indices, respectively.

The overall observation is that even though three different patterns are observed for the way *N* values change for different U.S. urban road networks, the parameters *a* and *b* obtained from the calibration of the fits continue to have similar meanings. Essentially, they represent the road density as well as the rate it drops while moving away from the center, therefore truly capturing their complex property. From the above three patterns, we see that they represent different rates of decay in the urban transportation networks, where the exponential fit is the fastest, followed by the power law and finishing with the logarithmic fit. Another difference is that the logarithmic trend, which is slow by nature, also exhibits an "additive" property, as opposed to the "multiplicative" nature of the other two fits.

Figure 2a illustrates the distributions of the "density index" within the three categories of cities. We can see that cities such as Minneapolis, Boston, and Chicago have denser centers as compared to cities such as Nashville, Charlotte, and Miami, respectively.

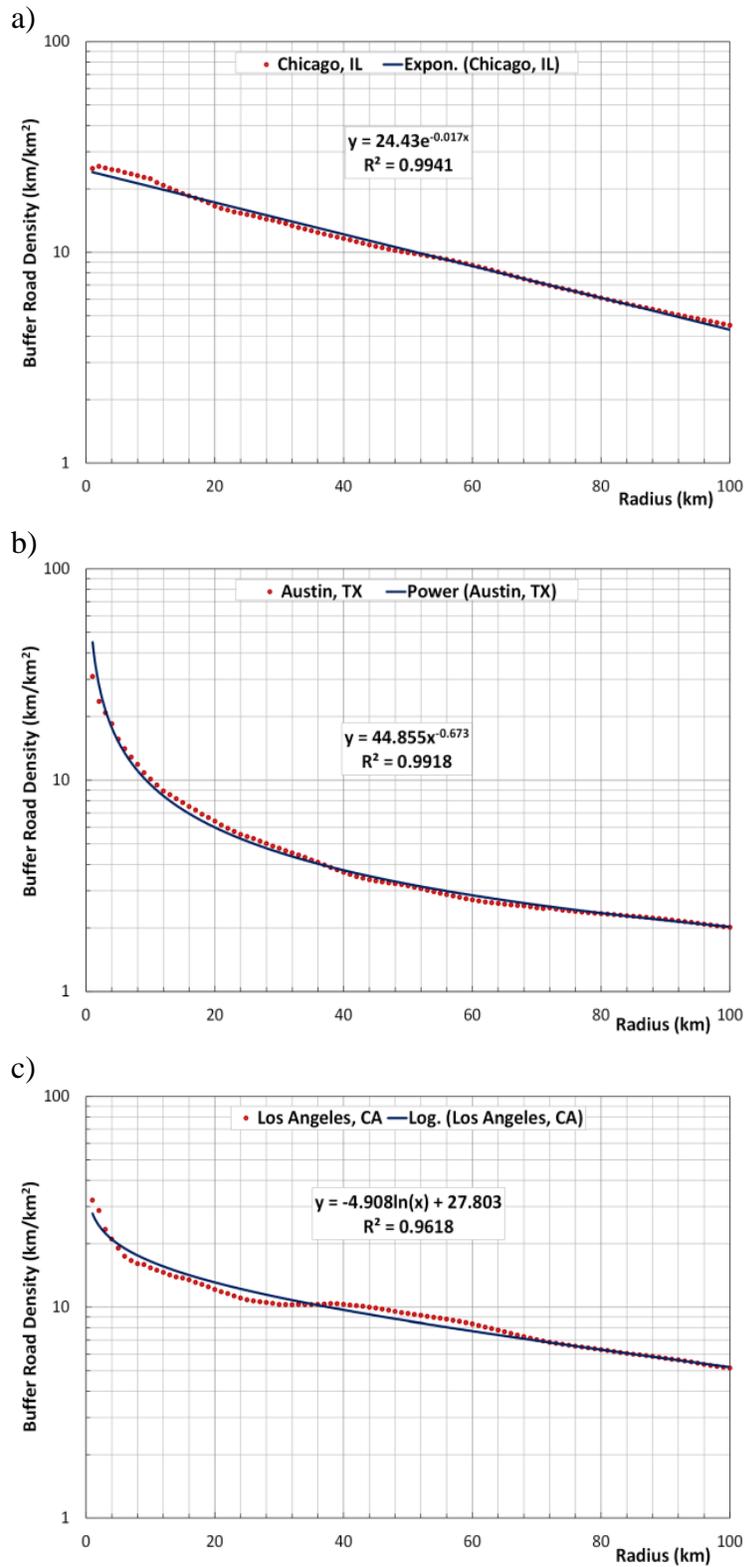

Figure 1 Semi-log plot of Buffer Road Density. a) Exponential fit in Chicago, IL. b) Power law fit in Austin, TX. c) Logarithmic fit in Los Angeles, CA.

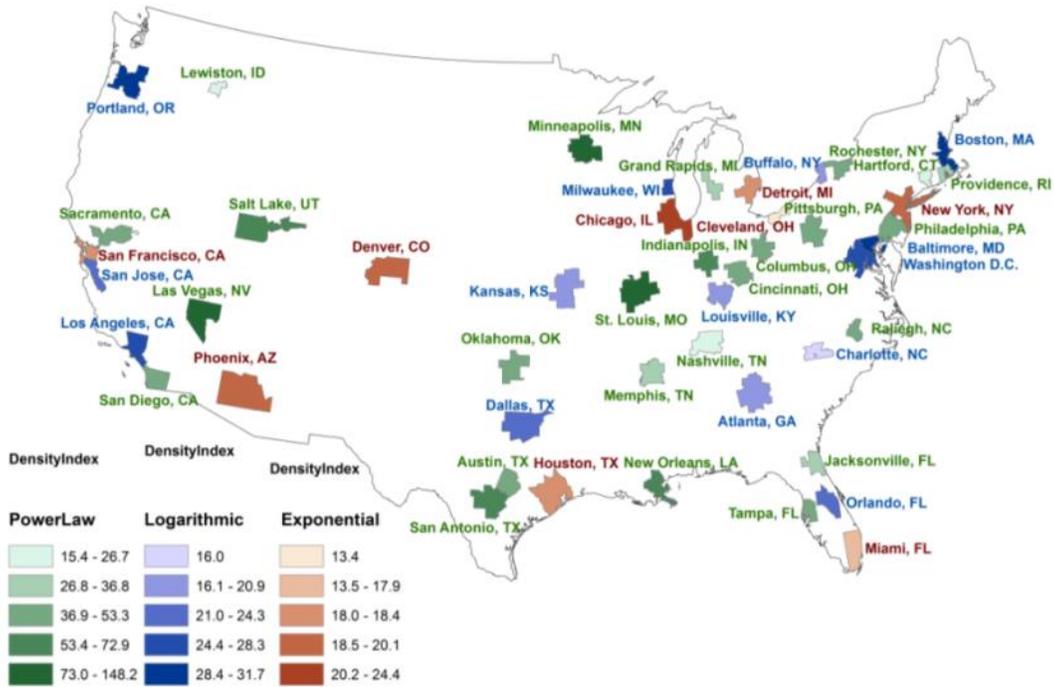

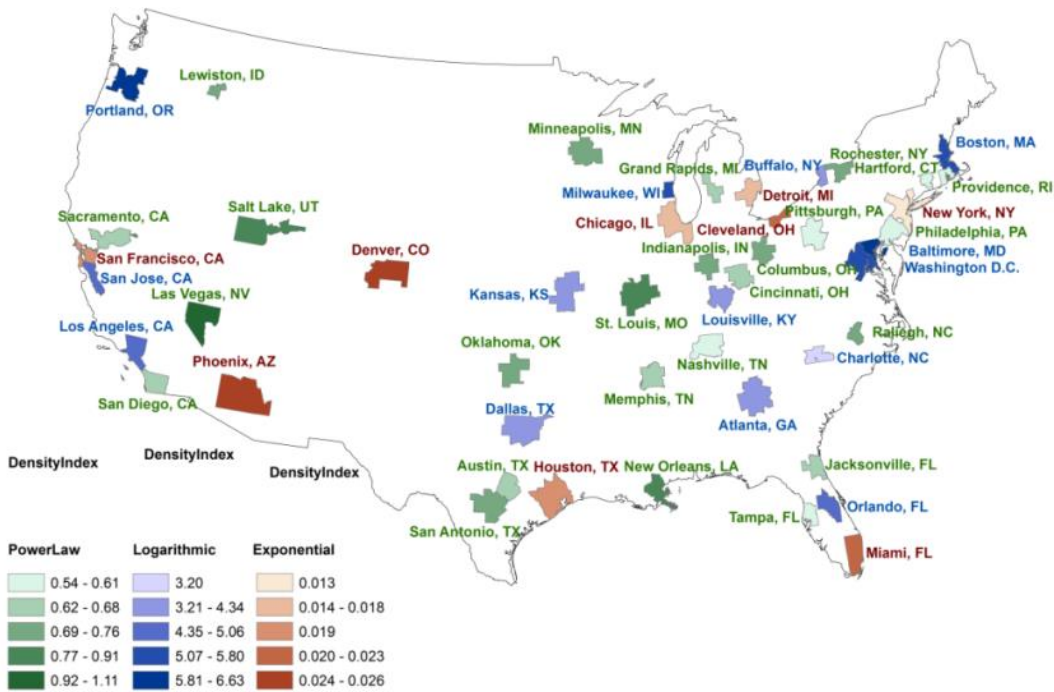

Figure 2 Spatial Distribution of the a) Density Index, and b) Decay Index.

(These maps were made in ArcGIS 10.2)

This observation agrees with the fact that the three former cities are indeed older and have initially developed in an era when walking was prevalent.

Moreover, Figure 2b demonstrates how the "decay index" varies among the same cities, based on which we see that in cities like Las Vegas, Portland, and Denver, the road networks' densities fall faster than in cities such as Sacramento, Buffalo, and San Francisco, respectively. In this case, the three former cities have relatively dense cores that rapidly evolve into less dense, suburban-type, road designs.

**DISCUSSION**

In the past, many studies have claimed that the transportation networks (as well as other components) of urban systems follow power law, and thus are fractals in nature [12,29,30,39]. In this study, under mono-centric assumption (described in the methods section), we developed a methodology for the application of a ring-buffer method as a tool for analyzing the coupled complexity of urban road networks. Even though the mathematical foundation of this method has been originally developed based on the assumption of the existence of a power law for the features being studied, we reject this assumption as a universal rule for urban road networks. Instead, we use a novel approach to decouple the mixed complex nature of urban transportation systems, through which we conclude that such features possess characteristics that are influenced by two components that are coupled. One is a power law with exponent 2 that captures the fractal aspect, or scaling property, of the road networks. The other component can come in three different forms, either exponential, or power law, or logarithmic, thus categorizing the urban road networks to three classes based on their complex nature and evolutionary path.

Based on these findings, two parameters are identified that can be systematically measured. One that has a direct and positive linear relationship with the total length and density of the road network, and because of that can be considered as a *compactness* or *density index*. And the other one that has a negative nonlinear inverse impact on the total length of the road network, and thus can be considered as a *decay index*. Using those two indices, various urban road networks can then be properly classified.

In short, through rigorous computational as well as analytical work, we show that regardless of the choice of the city, urban road networks possess similar attributes, while at the same time they also exhibit unique properties. In addition, our study rejects the universality of power law as the sole expression of the evolution of urban road networks, something that has been suggested by many researches in the past [12,29,30,39]. Instead, we show that urban road networks possess a combination of two characteristics; a scaling component related to the square of the radius, as well as a second component that can follow a number of trends (exponential, or logarithmic, or power law).

**METHODS**

**Description of the ring-buffer approach.** The ring-buffer method used in this study is based on the assumption that urban systems and their components, specifically their road networks, evolve similar to living organisms. A living being comes to life as a single cell. Then it grows and spreads around that center, subject to its prevailing conditions and constraints. Similar to that, a city spreads around a point of origin, or "center" [40,41], and then gradually expands outwards, while avoiding the physical constraints around it such as water bodies, etc. The widely accepted assumption is that the spread of any component of the urban system, e.g. its road network, at a given point is proportional to its distance from that center. Mathematically, for this assumption to hold true, it needs to manifest itself in the form of a power law. In other words, if measurements follow a power law, then the urban system, or its component, will be considered to be a *fractal*.

A fractal can be described as an entity that possesses self-similarity at all scales. It is important to note that a fractal needs to only exhibit similar (but not exactly the same) type of structure at all scales [42]. Moreover, according to Mandenlbrot: "A fractal set is one for which the fractal dimension strictly exceeds its topological dimension [42]." In practice, this means that while a line feature (e.g. a road) has a dimension of 1 in classical geometry, it must have a dimension larger than 1 (to a maximum of 2), if it is to have fractal properties.

The existence of a power law appears in the form of equation 5:

$$N(r) = a \cdot r^D \qquad (5)$$

in which $r$ is the radius (with respect to a point of origin or center), $N$ is the number quantifying the object under consideration within a circle of radius $r$, $a$ is a constant, and $D$ is the exponent, also called the *fractal dimension*. Figure 3a illustrates the idea, in which circles with increasing radii are created around the center. The quantities of the feature are calculated for each ring, and then successively added to obtain the quantities within concentric circles (or buffers) with the corresponding radii. The variation of the total length of the feature with respect to the buffer radius can then be examined for the presence of the power law, according to equation 5.

In order to facilitate the examination of the data, the measurements are typically plotted in log-log scale, as shown in Figure 3b. For that, taking the log of both sides in equation 5 results in:

$$Log[N(r)] = log(a) + D \cdot log(r) \tag{6}$$

in which $D$ (the fractal dimension) has become slope of the linear trend. We note that because the data points are ordered and successively plotted based on $r$ values, a regression analysis will be sufficient to linearly fit the outputs of this method to equation 6. The reader, however, is referred to [43] for a further discussion regarding statistical methods that can be used to fit power laws to overlapping data.

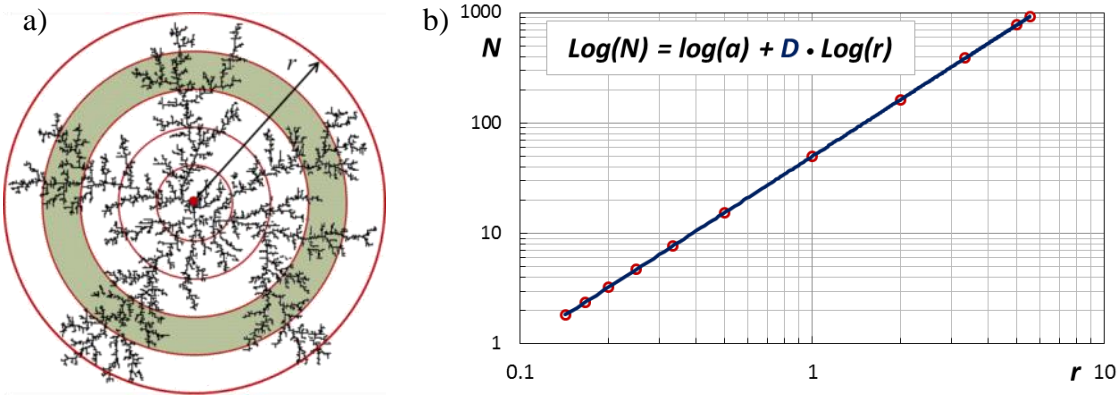

Figure 3 Ring-Buffer and Power Law. a) Ring creation in the ring-buffer method. b) Log-log plot of the power law relationship. The linear relationship in a log-log plot points to the presence of power law scaling property.

**Verification of Ring-Buffer Method for Fractal Analysis.** As the first step, the validity of the ring-buffer approach as a proper method for capturing the fractal nature of features is investigated. In order to do so, a Greek Cross grid, which is a well-known fractal with dimension of 2, is chosen. The rationale behind this choice is the resemblance of Greek Cross pattern to urban road systems, especially grid road networks. Also, to investigate if the grid cell size has any impact on the results, a total of 20 Greek Cross grids are created with varying cell sizes from 100 m to 10000 m. Moreover, and in order to capture the impact of boundary shapes on the results, the Greek Cross grids are all clipped by the Chicago MSA area. Figure 4a demonstrates one of the grids created via the above steps.

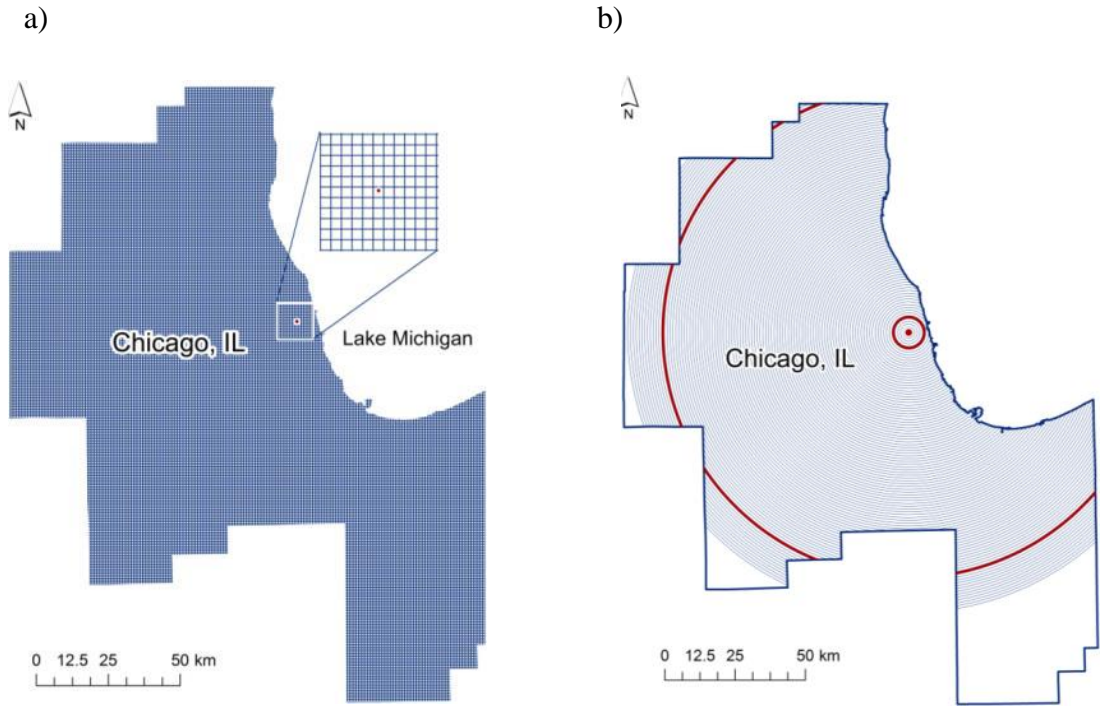

Figure 4 Greek Cross and Concentric Rings. a) Greek Cross grid with 1000 m cells created within Chicago MSA. b) Examples of full and partial rings.
(These maps were made in ArcGIS 10.2)

As we can see, there is no apparent center to the above grid network, mainly due to its uniform structure. As for the point around which the circles are to be created, therefore, the center chosen for the actual Chicago road network (as shown later in Figure 6) is used. The ring-buffer method is then applied to every grid created through the above steps, during which circles with the radii from 1 km to 100 km are created around the chosen center at the increments of 1 km. This results in the creation of a total of 100 rings of 1 km width, shown in Figure 4b.

The rings are then intersected with the grid networks, and the total road length within each ring is calculated for every grid. An important note to mention here is that at some radii, the boundary of Chicago MSA starts to cut through some of the rings, e.g., the largest red ring in Figure 4b, thus reducing the road lengths within the affected rings, as compared to the smaller rings that are uncut and complete. To rectify this problem, the density of the roads within the partial 1km rings are calculated and then extended to their corresponding full rings, as if no parts of them are cut. This allows us to successively add the ring road lengths to obtain the total road lengths within buffers (circles) around the center at the selected radii. The values obtained, which represent *N* in equation 5, are then plotted versus the radii in a log-log diagram, as shown in Figure 5.

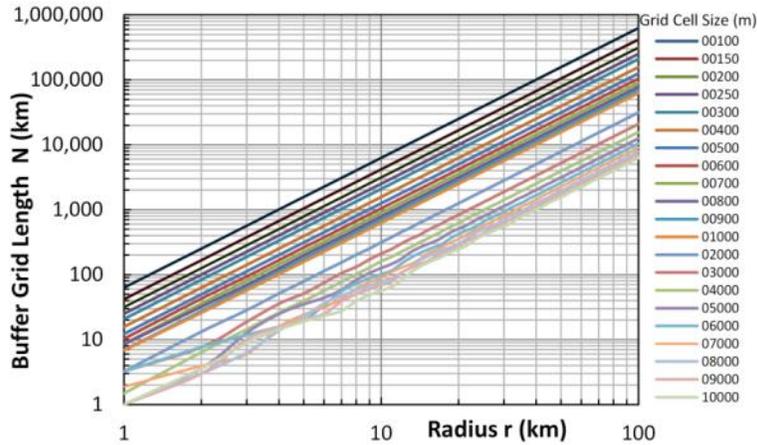

Figure 5 Log-log plots of buffer road length versus radius for different grid cell sizes.

Figure 5 shows that the fits to all the above plots follow linear patterns, i.e., they display power law relationships, supporting the existence of fractal properties. Moreover, the slopes of the fits to all the plots are equal to 2, meaning that the fractal dimensions of all the grids are 2, as expected. As it can be seen, for grids with large cell sizes (> 1000 m or 1 km), the plots show oscillations at the beginning, but still around a line. The reason is that the chosen ring width (1 km) becomes too small for grids with cell sizes of larger than 1 km. Nonetheless, all of the plots eventually become lines with the slope of 2. This investigation therefore validates the ability of the ring-buffer method to capture the characteristics of a fractal feature.

Another important observation is that the ring-buffer method is insensitive to the shape of the boundary of the chosen urban system, i.e. the shape and size of the MSA of a given urban system will not have an impact on the results. Moreover, the size of the grid cells used also does not affect the outcome of the ring-buffer method. Although the smaller the grid cell size is, the clearer the linear relationship becomes, even for larger grid sizes the oscillations remains around a line with the slope of 2, which therefore suggests that the choice of 1km as the ring width does not impact the conclusion.

**Application of Ring-Buffer Method to Urban Road Networks.** In order to investigate the complex properties of urban road networks, we apply this ring-buffer approach to the 50 U.S. urban systems listed in Appendix A. The first step is to select a consistent method for determining the "center" for any given road network. Based on the earlier discussion, we first use the distribution of the road density of the whole network over its MSA area to identify and select the densest area. Then, we choose the point with the highest road density within the selected area as the "center" for the whole network. An example of the application of this method to the Chicago MSA road network, and the selected center, is presented in Figure 6. Similar maps are generated for all 50 U.S. urban systems and are included in Appendix C.

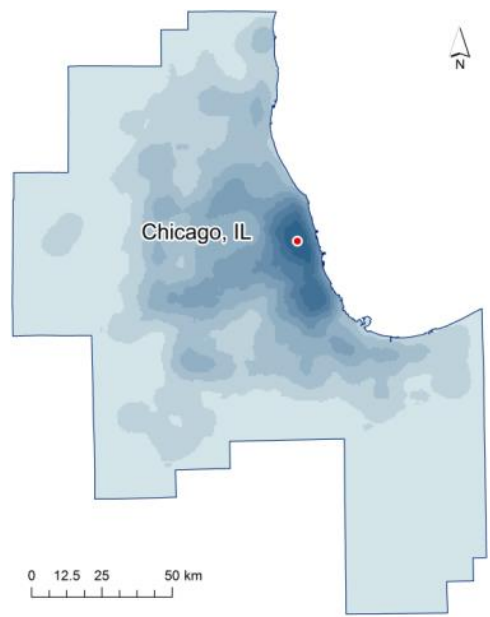

Figure 6 Road Density map for Chicago MSA road network, and the selected "center". (This map was made in ArcGIS 10.2)

Naturally, this method assumes a mono-centric urban form that shows a clear center. Although some cities have evolved to become poly-centric, our analysis shows that their road networks have often remained mono-centric, simply related to the fact that denser streets tend to locate in older areas of the cities. Even in a relatively young country as the U.S., only two cities out of the 50 cities studied did not have a clear center. In that case, a point between them is chosen as the center, for which the results are still found to be statistically significant (as shown in Appendix C).

The results of the application of the method developed in this study to the Chicago MSA road network are displayed in Figure 7a using a log-log scale. Looking purely at the data points (i.e., the blue dots), the trend looks close to linear, but the power law fit (i.e., red line) clearly shows a systematic bias and cannot be statistically validated. Another way to tackle the problem is to consider that if the original distribution is a power law, then dividing it by another power law should also result in a power law. This means that the density of roads, that we obtain by dividing buffer road length $N(r)$ by the area $A = \pi r^2$ (which is a power law), should also be a power law. Figure 7b, however, shows that the density as a function of radius clearly does not follow a power law relationship, which further points to the existence of another form that is yet to be explained.

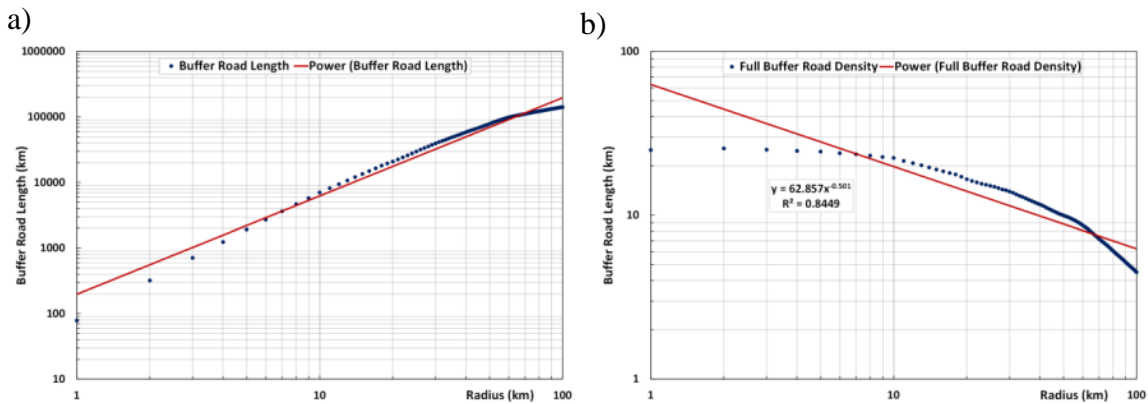

Figure 7 Road Length and Density in Chicago. a) Log-log plot of buffer road length versus radius. b) Log-log plot of buffer road density versus radius.

Similar steps are taken for all 50 U.S. urban areas chosen for the study. A significant number of plots also show the same issue. Therefore, we must reject the universality of power law as the manifestation of the evolution of urban road networks. Instead, we decouple the complexity of urban road networks as the product of area and road density, as expressed in equation 1, which offers us statistically significant fits, examples of which can be seen in Figure 1 and also Appendix C.

# SUPPLEMENTARY MATERIALS

# APPENDIX A

**List of 50 U.S. Urban Systems Studied**

Table A1 List of 50 U.S. urban systems studied

| Urban Area, State | Founded in[1] | Population[2] | Area (km$^2$)[3] | Pop Density | Road Length (km)[3] | # of Intersections[3] |
|---|---|---|---|---|---|---|
| Atlanta, GA | 1843 | 5486738 | 20306.8 | 270.2 | 67215.1 | 243462 |
| Austin, TX | 1835 | 1784094 | 9440.8 | 189.0 | 30382.0 | 111234 |
| Baltimore, MD | 1729 | 2895944 | 5624.9 | 514.8 | 35556.3 | 220784 |
| Boston, MA | 1630 | 4892136 | 8368.7 | 584.6 | 49139.9 | 261949 |
| Buffalo, NY | 1789 | 1191744 | 3821.4 | 311.9 | 12293.0 | 41429 |
| Carson, NV | 1858 | 87743 | 109.0 | 804.9 | 900.6 | 3045 |
| Charlotte, NC | 1755 | 1927130 | 7177.5 | 268.5 | 24978.8 | 93988 |
| Chicago, IL | 1803 | 9594379 | 17783.6 | 539.5 | 86788.9 | 396704 |
| Cincinnati, OH | 1788 | 2252951 | 10398.8 | 216.7 | 33834.5 | 141744 |
| Cleveland, OH | 1796 | 2272776 | 4827.5 | 470.8 | 19472.2 | 64630 |
| Columbus, OH | 1812 | 1949603 | 9483.2 | 205.6 | 27764.3 | 106156 |
| Dallas, TX | 1841 | 6501589 | 21833.1 | 297.8 | 83815.2 | 350762 |
| Denver, CO | 1858 | 2666592 | 18262.0 | 146.0 | 46547.0 | 182157 |
| Detroit, MI | 1701 | 4369224 | 9664.6 | 452.1 | 46880.4 | 187960 |
| Grand Rapids, MI | 1825 | 895227 | 6665.8 | 134.3 | 16684.6 | 42990 |
| Hartford, CT | 1637 | 1400709 | 3487.6 | 401.6 | 14992.7 | 56695 |
| Honolulu, HI | 1809 | 953207 | 775.4 | 1229.3 | 4678.9 | 22904 |

Table A1 List of 50 U.S. urban systems studied

| Urban Area, State | Founded in[1] | Population[2] | Area (km$^2$)[3] | Pop Density | Road Length (km)[3] | # of Intersections[3] |
|---|---|---|---|---|---|---|
| Houston, TX | 1837 | 6052475 | 20585.7 | 294.0 | 83365.0 | 353831 |
| Indianapolis, IN | 1821 | 1856996 | 9289.1 | 199.9 | 32389.9 | 150469 |
| Jacksonville, FL | 1822 | 1451740 | 7182.3 | 202.1 | 22067.4 | 76396 |
| Kansas City, KS | 1868 | 2138010 | 19148.1 | 111.7 | 50639.6 | 184748 |
| Las Vegas, NV | 1905 | 2010951 | 7330.1 | 274.3 | 20926.8 | 104925 |
| Lewiston, ID | 1861 | 85096 | 2104.6 | 40.4 | 4206.1 | 6334 |
| Los Angeles, CA | 1781 | 13059105 | 10913.2 | 1196.6 | 70096.7 | 335638 |
| Louisville, KY | 1778 | 1443801 | 9227.8 | 156.5 | 24453.7 | 82680 |
| Memphis, TN | 1819 | 1398172 | 10049.2 | 139.1 | 25028.4 | 74462 |
| Miami, FL | 1896 | 5571523 | 8410.3 | 662.5 | 42827.1 | 178680 |
| Milwaukee, WI | 1833 | 1602022 | 3507.8 | 456.7 | 17207.1 | 66802 |
| Minneapolis, MN | 1867 | 3412291 | 15365.8 | 222.1 | 57532.0 | 259788 |
| Nashville, TN | 1779 | 1740134 | 13588.3 | 128.1 | 32653.8 | 90700 |
| New Orleans, LA | 1718 | 1247062 | 3715.5 | 335.6 | 18340.7 | 83361 |
| New York, NY | 1624 | 19217139 | 15551.5 | 1235.7 | 105344.0 | 499969 |
| Oklahoma, OK | 1889 | 1359027 | 13051.0 | 104.1 | 34167.6 | 120303 |
| Orlando, FL | 1875 | 2257901 | 7996.6 | 282.4 | 28876.5 | 123076 |

Table A1 List of 50 U.S. urban systems studied

| Urban Area, State | Founded in[1] | Population[2] | Area (km$^2$)[3] | Pop Density | Road Length (km)[3] | # of Intersections[3] |
|---|---|---|---|---|---|---|
| Philadelphia, PA | 1682 | 6234336 | 11271.7 | 553.1 | 58104.3 | 256023 |
| Phoenix, AZ | 1868 | 4262838 | 25763.0 | 165.5 | 60738.6 | 241836 |
| Pittsburgh, PA | 1758 | 2503836 | 12859.9 | 194.7 | 45196.4 | 167027 |
| Portland, OR | 1845 | 2363554 | 14669.4 | 161.1 | 44544.0 | 174765 |
| Providence, RI | 1636 | 1695760 | 3773.5 | 449.4 | 18431.5 | 83871 |
| Raliegh, NC | 1792 | 1258825 | 4830.5 | 260.6 | 18678.0 | 81802 |
| Rochester, NY | 1803 | 1159166 | 7037.2 | 164.7 | 17863.9 | 47275 |
| Sacramento, CA | 1839 | 2277843 | 10167.0 | 224.0 | 34020.6 | 124839 |
| Salt Lake, UT | 1847 | 1246208 | 10895.1 | 114.4 | 22387.0 | 59736 |
| San Antonio, TX | 1718 | 2239307 | 16213.5 | 138.1 | 44137.5 | 127773 |
| San Diego, CA | 1769 | 3144425 | 7668.0 | 410.1 | 29499.1 | 144194 |
| San Francisco, CA | 1776 | 4472992 | 5352.1 | 835.7 | 33483.0 | 172400 |
| San Jose, CA | 1777 | 1992872 | 4921.2 | 405.0 | 19824.6 | 93610 |
| St. Louis, MO | 1763 | 2934412 | 20184.1 | 145.4 | 57670.8 | 205269 |
| Tampa, FL | 1823 | 2858974 | 5756.8 | 496.6 | 31421.2 | 143714 |
| Washington D.C. | 1790 | 5916033 | 12735.0 | 464.5 | 74190.6 | 437470 |

1. Wikipedia, Accessed 2014-06: http://www.wikipedia.org/
2. U.S. Census Bureau American FactFinder, 2010: http://factfinder2.census.gov/
3. Calculated from U.S. Census Bureau TIGER/Line Shapefiles, 2010: https://www.census.gov/geo/maps-data/data/tiger-line.html

# APPENDIX B

**Density and Decay Indices for 50 U.S. Urban Road Networks**

Table B1 Density and Decay Indices for 50 U.S. Urban Road Networks

| Urban Area, State | Density Index (km) | Decay Index (1/km) | Fit Type |
|---|---:|---:|---|
| Atlanta, GA | 20.24 | 3.9 | Logarithmic |
| Austin, TX | 44.855 | 0.673 | Power Law |
| Baltimore, MD | 31.692 | 6.633 | Logarithmic |
| Boston, MA | 29.722 | 5.79 | Logarithmic |
| Buffalo, NY | 19.897 | 4.344 | Logarithmic |
| Carson, NV | 22.281 | 0.914 | Power Law |
| Charlotte, NC | 16.034 | 3.198 | Logarithmic |
| Chicago, IL | 24.43 | 0.017 | Exponential |
| Cincinnati, OH | 49.53 | 0.675 | Power Law |
| Cleveland, OH | 13.39 | 0.023 | Exponential |
| Columbus, OH | 47.732 | 0.692 | Power Law |
| Dallas, TX | 22.61 | 4.244 | Logarithmic |
| Denver, CO | 19.475 | 0.026 | Exponential |
| Detroit, MI | 18.225 | 0.018 | Exponential |
| Grand Rapids, MI | 30.192 | 0.632 | Power Law |
| Hartford, CT | 26.681 | 0.541 | Power Law |
| Honolulu, HI | 22.073 | 0.586 | Power Law |
| Houston, TX | 18.448 | 0.019 | Exponential |
| Indianapolis, IN | 72.927 | 0.755 | Power Law |
| Jacksonville, FL | 32.15 | 0.625 | Power Law |
| Kansas City, KS | 20.921 | 4.207 | Logarithmic |
| Las Vegas, NV | 148.2 | 1.105 | Power Law |
| Lewiston, ID | 15.368 | 0.721 | Power Law |
| Los Angeles, CA | 27.803 | 4.908 | Logarithmic |
| Louisville, KY | 18.781 | 4.192 | Logarithmic |

Table B1 Density and Decay Indices for 50 U.S. Urban Road Networks

| Urban Area, State | Density Index (km) | Decay Index (1/km) | Fit Type |
|---|---:|---:|---|
| Memphis, TN | 32.929 | 0.656 | Power Law |
| Miami, FL | 17.908 | 0.021 | Exponential |
| Milwaukee, WI | 27.063 | 5.803 | Logarithmic |
| Minneapolis, MN | 92.52 | 0.742 | Power Law |
| Nashville, TN | 26.233 | 0.576 | Power Law |
| New Orleans, LA | 63.23 | 0.854 | Power Law |
| New York, NY | 20.149 | 0.013 | Exponential |
| Oklahoma, OK | 51.494 | 0.712 | Power Law |
| Orlando, FL | 21.955 | 4.632 | Logarithmic |
| Philadelphia, PA | 52.557 | 0.571 | Power Law |
| Phoenix, AZ | 19.883 | 0.026 | Exponential |
| Pittsburgh, PA | 43.931 | 0.606 | Power Law |
| Portland, OR | 29.404 | 6.371 | Logarithmic |
| Providence, RI | 36.808 | 0.586 | Power Law |
| Raliegh, NC | 53.298 | 0.723 | Power Law |
| Rochester, NY | 40.877 | 0.709 | Power Law |
| Sacramento, CA | 42.276 | 0.643 | Power Law |
| Salt Lake, UT | 55.928 | 0.818 | Power Law |
| San Antonio, TX | 59.363 | 0.725 | Power Law |
| San Diego, CA | 44.855 | 0.666 | Power Law |
| San Francisco, CA | 18.316 | 0.019 | Exponential |
| San Jose, CA | 24.289 | 5.063 | Logarithmic |
| St. Louis, MO | 92.722 | 0.801 | Power Law |
| Tampa, FL | 46.73 | 0.576 | Power Law |
| Washington D.C. | 28.348 | 5.341 | Logarithmic |

**APPENDIX C**

**Characteristic Maps for 50 U.S. Urban Systems**

# Atlanta, GA

Road Network

Road Polygon Area

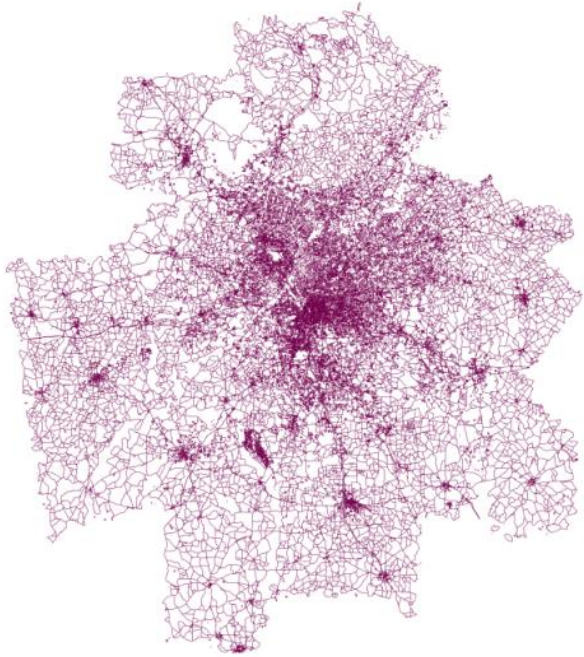

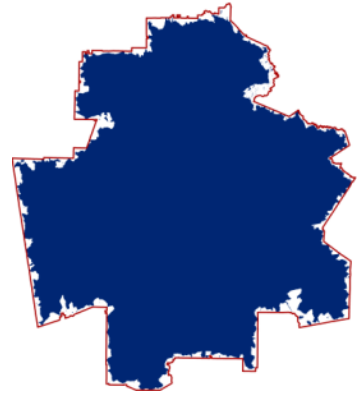

Road Density Map

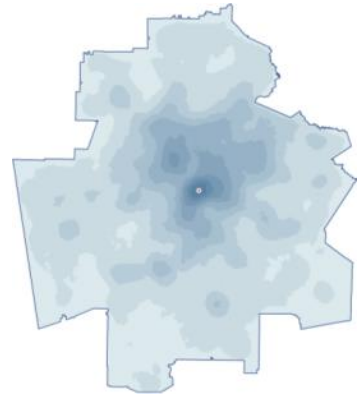

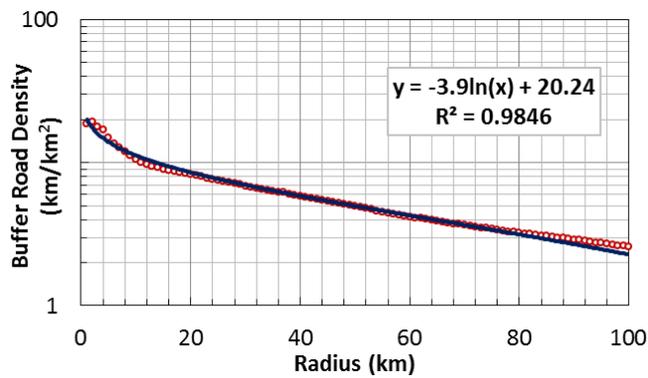

Road Density Fit

### Characteristics

| | |
|---|---:|
| Founded in | 1843 |
| Population | 5486738 |
| Pop Density | 270.2 |
| Area (km$^2$) | 20306.8 |
| Road Length (km) | 67215.1 |
| # of Intersections | 243462 |
| Area Threshold | 872 |
| Line Threshold | 610 |
| Point Threshold | 270 |
| Density Index | 20.24 |
| Decay Index | 3.9 |

# Austin, TX

Road Network

Road Polygon Area

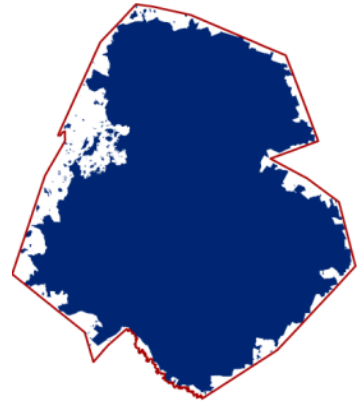

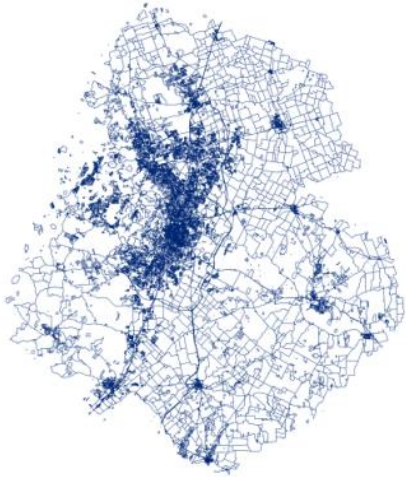

Road Density Map

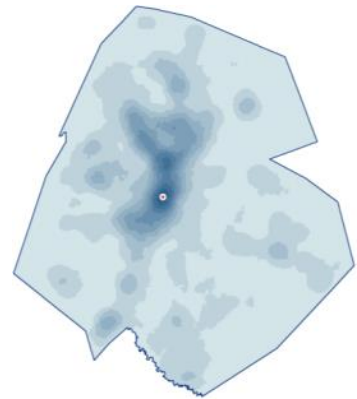

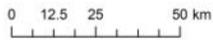
Austin, TX

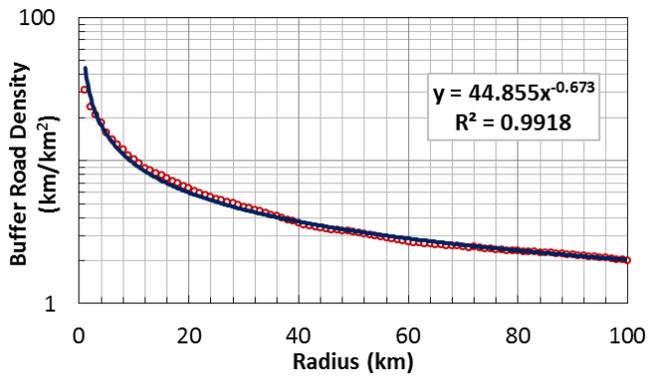

Road Density Fit

## Characteristics

| | |
|---|---:|
| Founded in | 1835 |
| Population | 1784094 |
| Pop Density | 189 |
| Area (km²) | 9440.8 |
| Road Length (km) | 30382 |
| # of Intersections | 111234 |
| Area Threshold | 794 |
| Line Threshold | 666 |
| Point Threshold | 272 |
| Density Index | 44.855 |
| Decay Index | 0.673 |

# Baltimore, MD

Road Network

Road Polygon Area

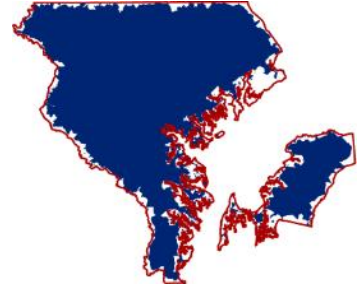

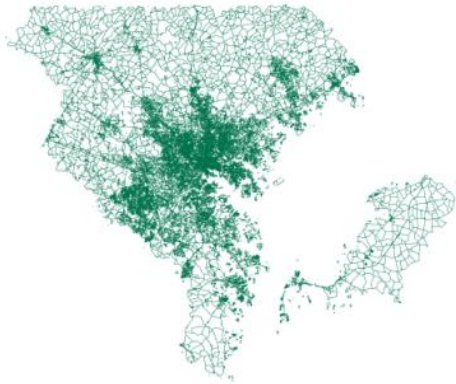

Road Density Map

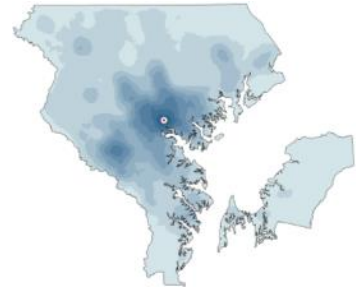

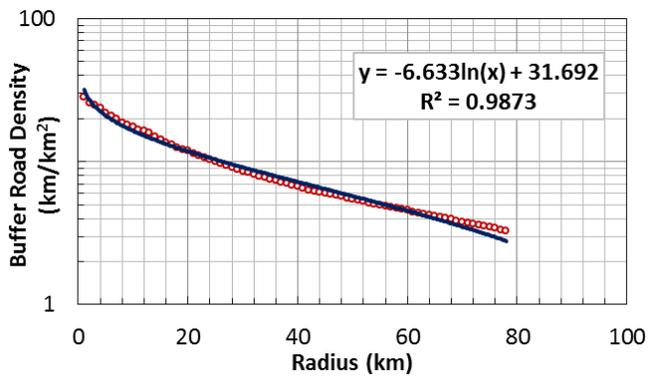

Road Density Fit

## Characteristics

| | |
|---|---:|
| Founded in | 1729 |
| Population | 2895944 |
| Pop Density (/km$^2$) | 514.8 |
| Area (km$^2$) | 5624.9 |
| Road Length (km) | 35556.3 |
| # of Intersections | 220784 |
| Area Threshold (m) | 390 |
| Line Threshold (m) | 352 |
| Point Threshold (m) | 153 |
| Density Index (km$^2$) | 31.692 |
| Decay Index (1/km) | 6.633 |

# **Boston, MA**

Road Network

Road Polygon Area

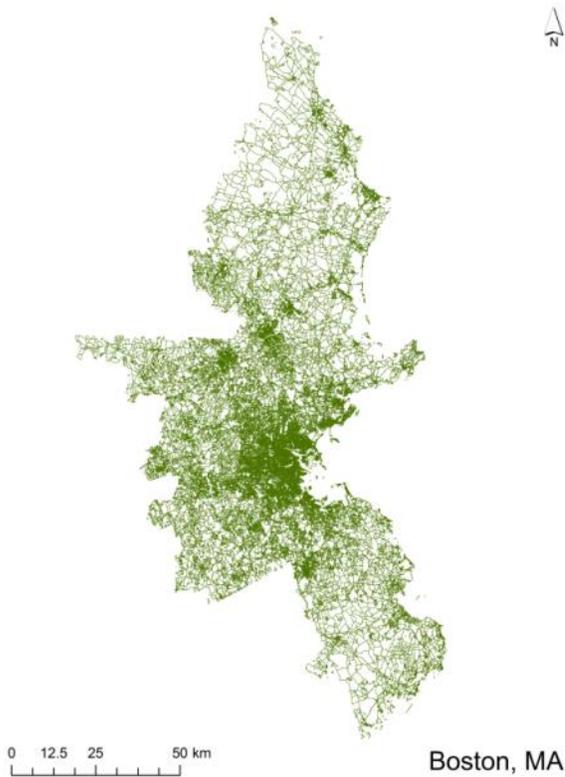

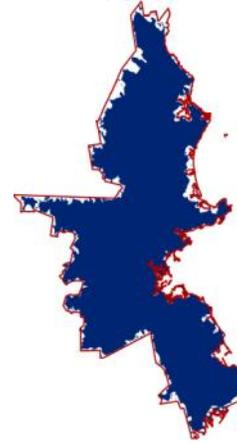

Road Density Map

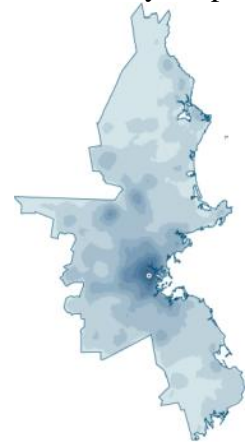

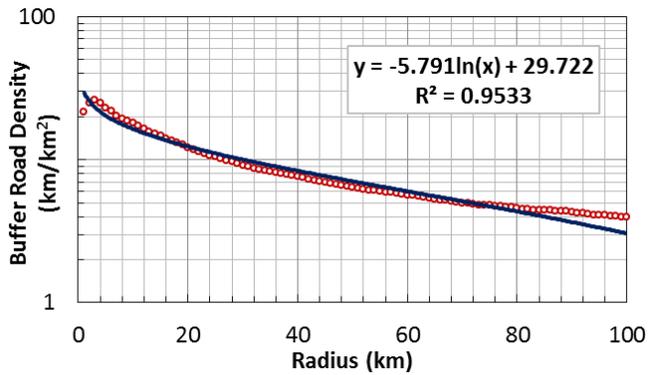

Road Density Fit

### **Characteristics**

| | |
|---|---:|
| Founded in | 1630 |
| Population | 4892136 |
| Pop Density (/km$^2$) | 584.6 |
| Area (km$^2$) | 8368.7 |
| Road Length (km) | 49139.9 |
| # of Intersections | 261949 |
| Area Threshold (m) | 480 |
| Line Threshold (m) | 353 |
| Point Threshold (m) | 174 |
| Density Index (km$^2$) | 29.722 |
| Decay Index (1/km) | 5.79 |

# **Buffalo, NY**

Road Network

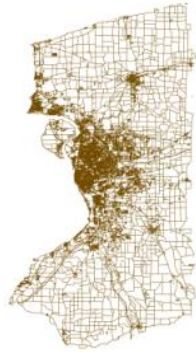

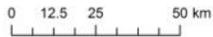
Buffalo, NY

Road Polygon Area

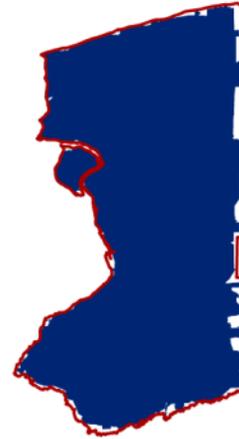

Road Density Map

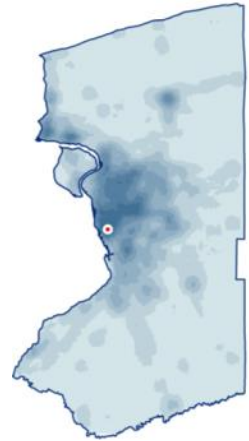

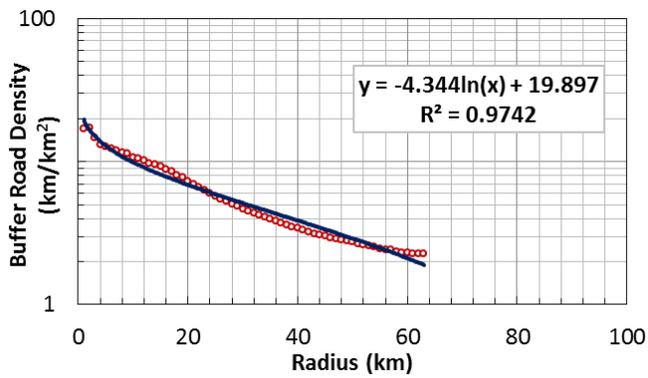

Road Density Fit

### **Characteristics**

| | |
|---|---:|
| Founded in | 1789 |
| Population | 1191744 |
| Pop Density (/km$^2$) | 311.9 |
| Area (km$^2$) | 3821.4 |
| Road Length (km) | 12293 |
| # of Intersections | 41429 |
| Area Threshold (m) | 971 |
| Line Threshold (m) | 615 |
| Point Threshold (m) | 270 |
| Density Index (km$^2$) | 19.897 |
| Decay Index (1/km) | 4.344 |

# Carson, NV

Road Network                                            Road Polygon Area

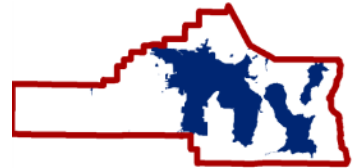

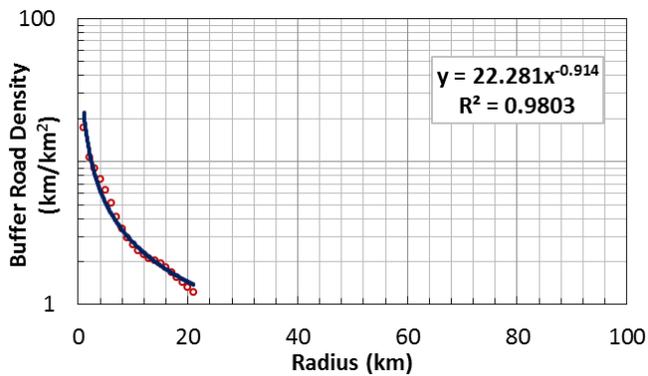

Road Density Map

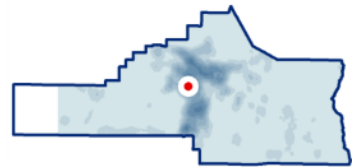

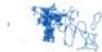

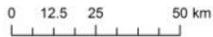

Carson, NV

| Characteristics | |
|---|---:|
| Founded in | 1858 |
| Population | 87743 |
| Pop Density (/km$^2$) | 804.9 |
| Area (km$^2$) | 109 |
| Road Length (km) | 900.6 |
| # of Intersections | 3045 |
| Area Threshold (m) | 156 |
| Line Threshold (m) | 778 |
| Point Threshold (m) | 279 |
| Density Index (km$^2$) | 22.281 |
| Decay Index (1/km) | 0.914 |

Road Density Fit

$y = 22.281x^{-0.914}$
$R^2 = 0.9803$

# Charlotte, NC

Road Network

Road Polygon Area

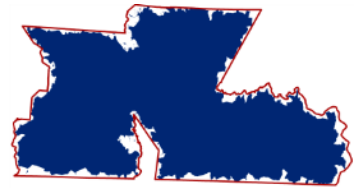

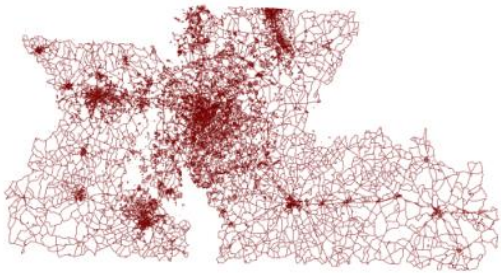

Road Density Map

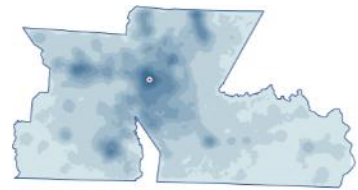

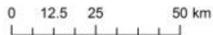

Charlotte, NC

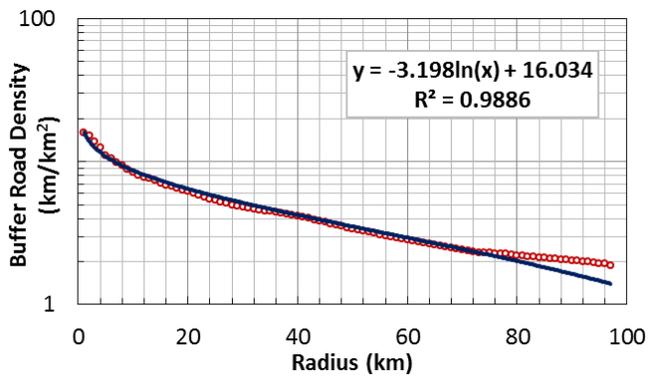

Road Density Fit

### Characteristics

| | |
|---|---:|
| Founded in | 1755 |
| Population | 1927130 |
| Pop Density (/km$^2$) | 268.5 |
| Area (km$^2$) | 7177.5 |
| Road Length (km) | 24978.8 |
| # of Intersections | 93988 |
| Area Threshold (m) | 672 |
| Line Threshold (m) | 627 |
| Point Threshold (m) | 267 |
| Density Index (km$^2$) | 16.034 |
| Decay Index (1/km) | 3.198 |

# Chicago, IL

Road Network

Road Polygon Area

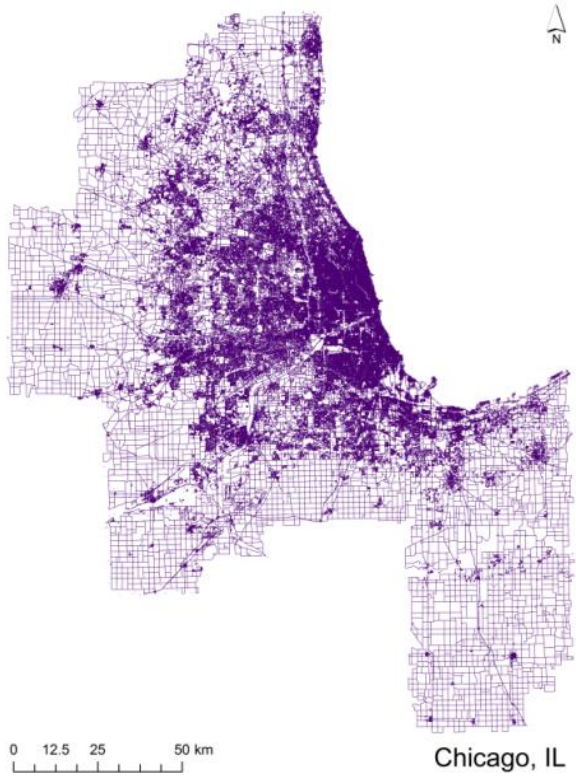

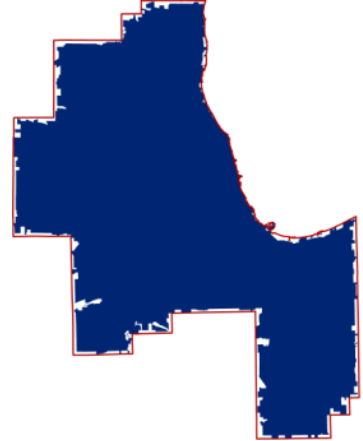

Road Density Map

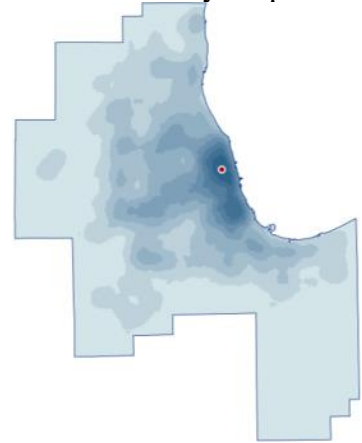

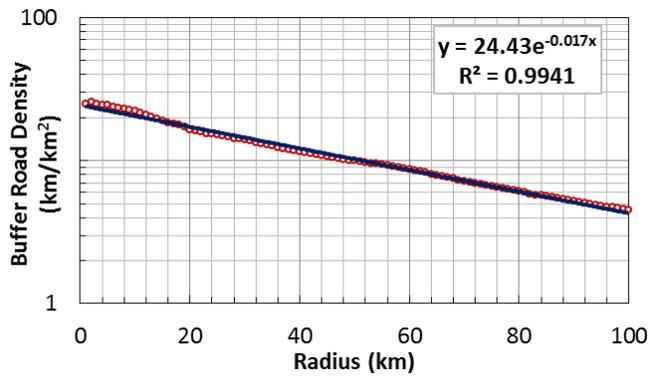

Road Density Fit

### Characteristics

| | |
|---|---:|
| Founded in | 1803 |
| Population | 9594379 |
| Pop Density (/km$^2$) | 539.5 |
| Area (km$^2$) | 17783.6 |
| Road Length (km) | 86788.9 |
| # of Intersections | 396704 |
| Area Threshold (m) | 984 |
| Line Threshold (m) | 321 |
| Point Threshold (m) | 179 |
| Density Index (km$^2$) | 24.43 |
| Decay Index (1/km) | 0.017 |

# Cincinnati, OH

Road Network

Road Polygon Area

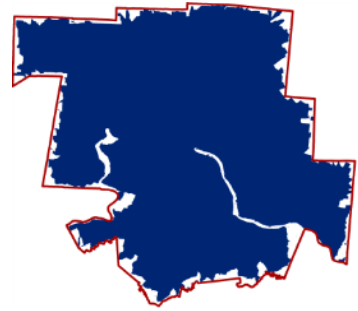

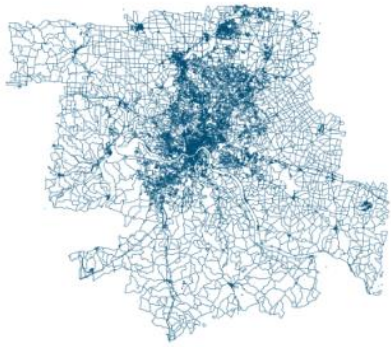

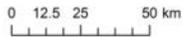

Cincinnati, OH

Road Density Map

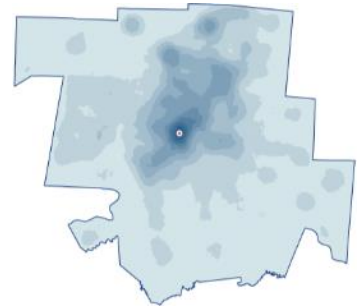

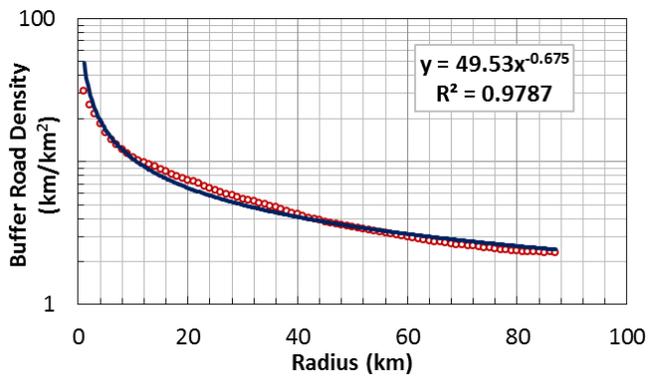

Road Density Fit

## Characteristics

| | |
|---|---:|
| Founded in | 1788 |
| Population | 2252951 |
| Pop Density (/km$^2$) | 216.7 |
| Area (km$^2$) | 10398.8 |
| Road Length (km) | 33834.5 |
| # of Intersections | 141744 |
| Area Threshold (m) | 745 |
| Line Threshold (m) | 668 |
| Point Threshold (m) | 249 |
| Density Index (km$^2$) | 49.53 |
| Decay Index (1/km) | 0.675 |

$y = 49.53x^{-0.675}$
$R^2 = 0.9787$

# Cleveland, OH

Road Network

Road Polygon Area

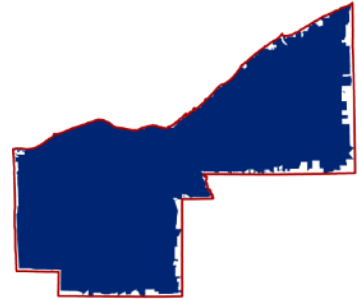

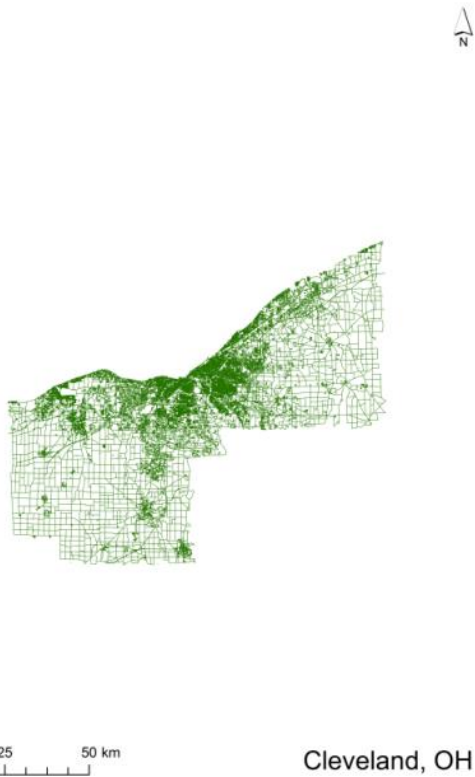

Road Density Map

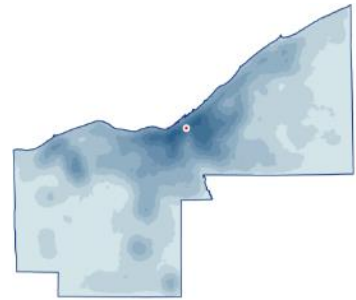

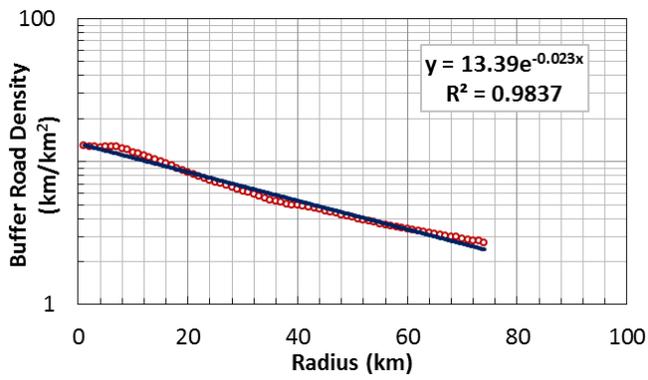

Road Density Fit

### Characteristics

| | |
|---|---:|
| Founded in | 1796 |
| Population | 2272776 |
| Pop Density (/km$^2$) | 470.8 |
| Area (km$^2$) | 4827.5 |
| Road Length (km) | 19472.2 |
| # of Intersections | 64630 |
| Area Threshold (m) | 821 |
| Line Threshold (m) | 479 |
| Point Threshold (m) | 251 |
| Density Index (km$^2$) | 13.39 |
| Decay Index (1/km) | 0.023 |

Fit equation: $y = 13.39e^{-0.023x}$, $R^2 = 0.9837$

# Columbus, OH

Road Network

Road Polygon Area

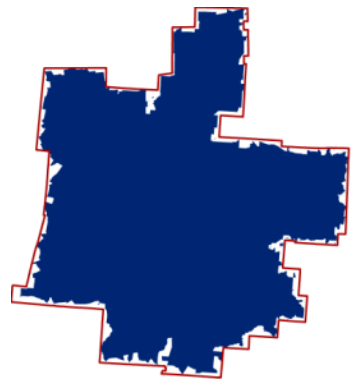

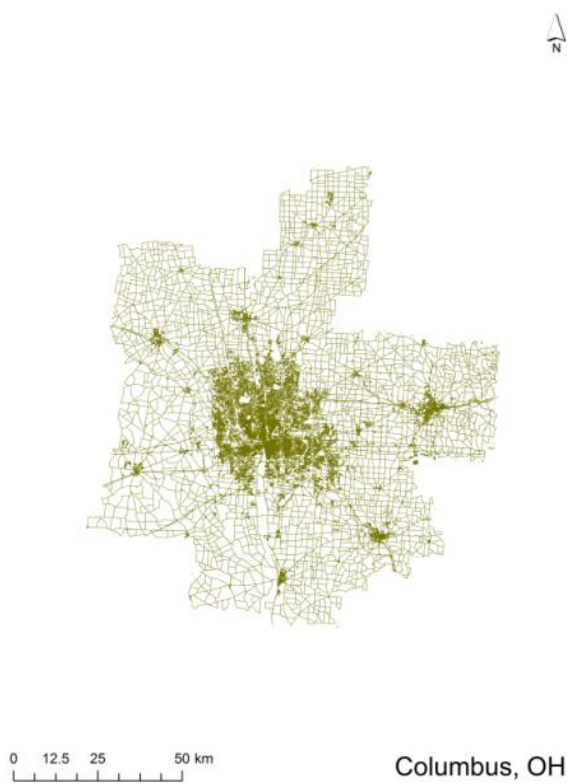

Road Density Map

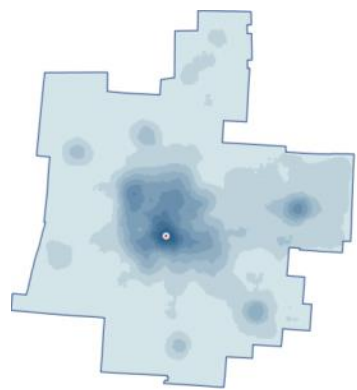

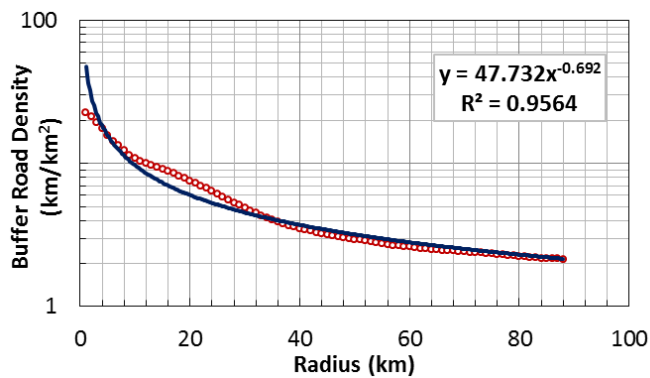

Road Density Fit

## Characteristics

| | |
|---|---:|
| Founded in | 1812 |
| Population | 1949603 |
| Pop Density (/km$^2$) | 205.6 |
| Area (km$^2$) | 9483.2 |
| Road Length (km) | 27764.3 |
| # of Intersections | 106156 |
| Area Threshold (m) | 907 |
| Line Threshold (m) | 722 |
| Point Threshold (m) | 267 |
| Density Index (km$^2$) | 47.732 |
| Decay Index (1/km) | 0.692 |

# Dallas, TX

Road Network

Road Polygon Area

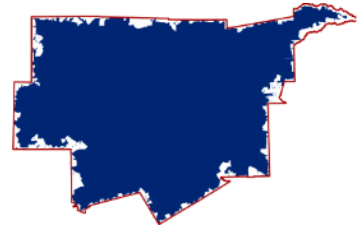

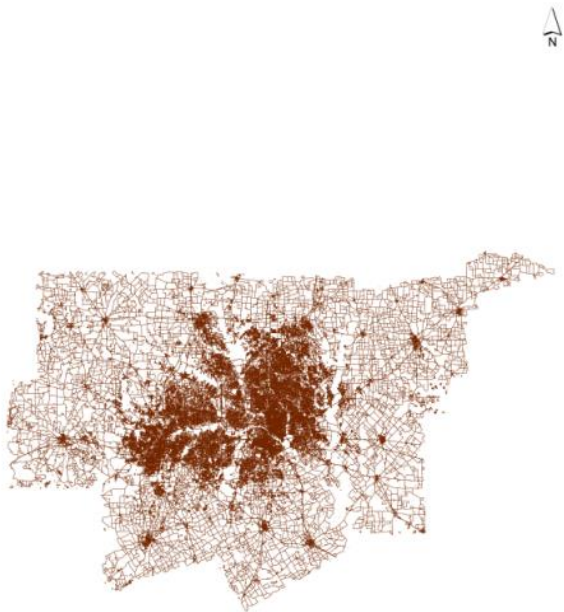

Road Density Map

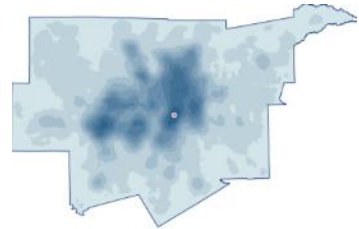

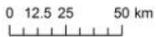
Dallas, TX

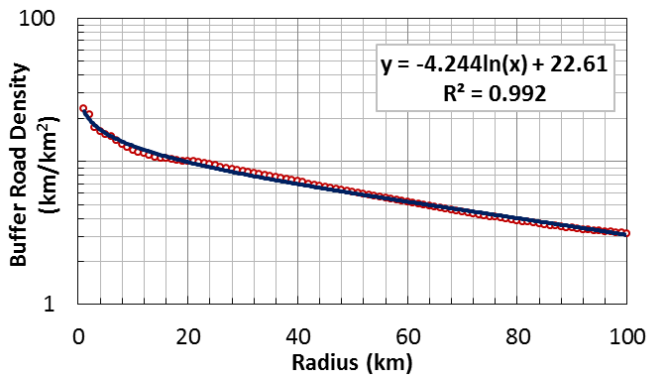

Road Density Fit

**Characteristics**

| | |
|---|---:|
| Founded in | 1841 |
| Population | 6501589 |
| Pop Density (/km$^2$) | 297.8 |
| Area (km$^2$) | 21833.1 |
| Road Length (km) | 83815.2 |
| # of Intersections | 350762 |
| Area Threshold (m) | 971 |
| Line Threshold (m) | 472 |
| Point Threshold (m) | 215 |
| Density Index (km$^2$) | 22.61 |
| Decay Index (1/km) | 4.244 |

Fit equation: $y = -4.244\ln(x) + 22.61$, $R^2 = 0.992$

# **Denver, CO**

Road Network

Road Polygon Area

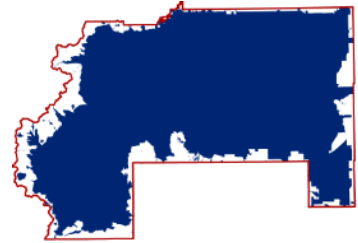

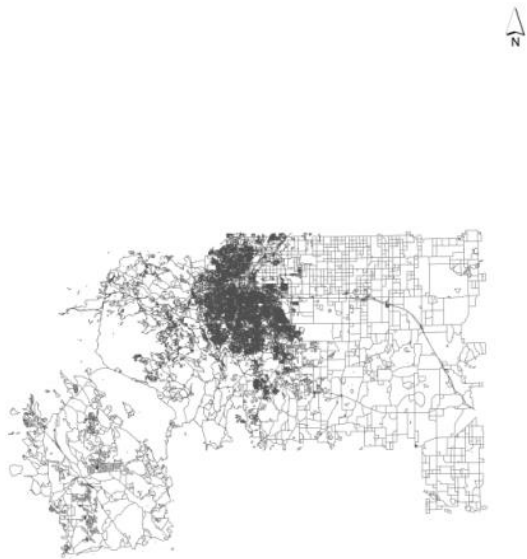

Road Density Map

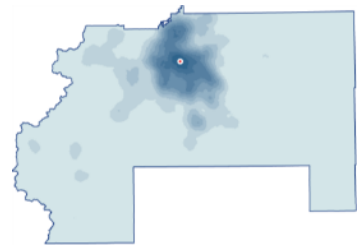

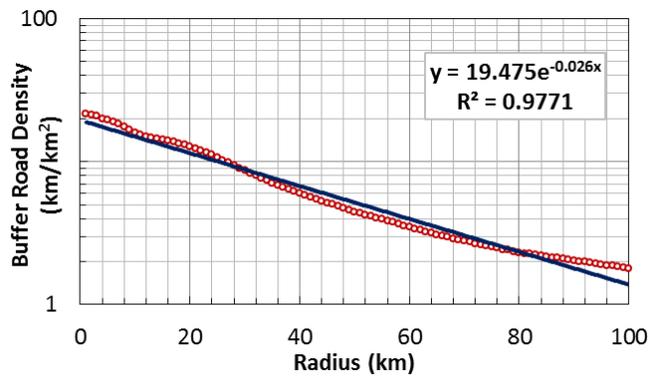

Road Density Fit

### **Characteristics**

| | |
|---|---:|
| Founded in | 1858 |
| Population | 2666592 |
| Pop Density (/km$^2$) | 146 |
| Area (km$^2$) | 18262 |
| Road Length (km) | 46547 |
| # of Intersections | 182157 |
| Area Threshold (m) | 1671 |
| Line Threshold (m) | 654 |
| Point Threshold (m) | 241 |
| Density Index (km$^2$) | 19.475 |
| Decay Index (1/km) | 0.026 |

# **Detroit, MI**

Road Network                                         Road Polygon Area

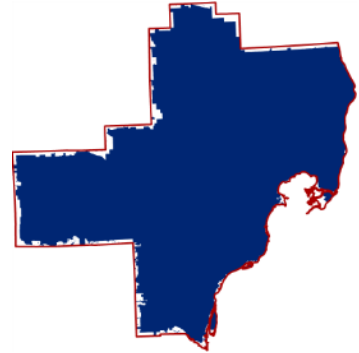

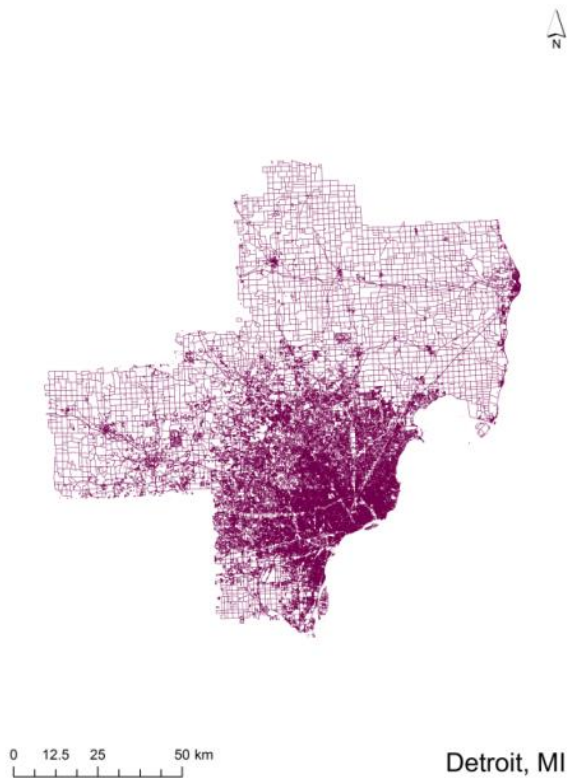

Road Density Map

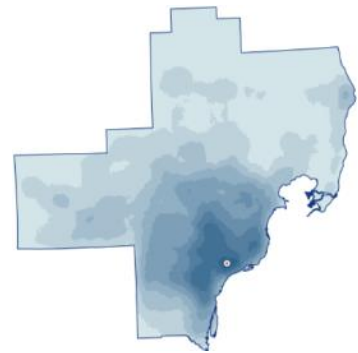

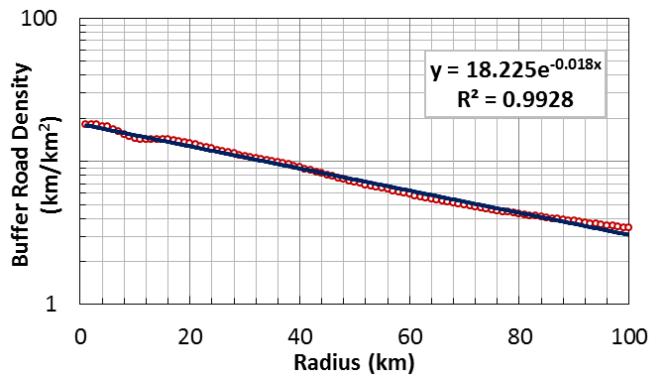

Road Density Fit

### **Characteristics**

| | |
|---|---:|
| Founded in | 1701 |
| Population | 4369224 |
| Pop Density (/km$^2$) | 452.1 |
| Area (km$^2$) | 9664.6 |
| Road Length (km) | 46880.4 |
| # of Intersections | 187960 |
| Area Threshold (m) | 751 |
| Line Threshold (m) | 381 |
| Point Threshold (m) | 204 |
| Density Index (km$^2$) | 18.225 |
| Decay Index (1/km) | 0.018 |

# Grand Rapids, MI

Road Network

Road Polygon Area

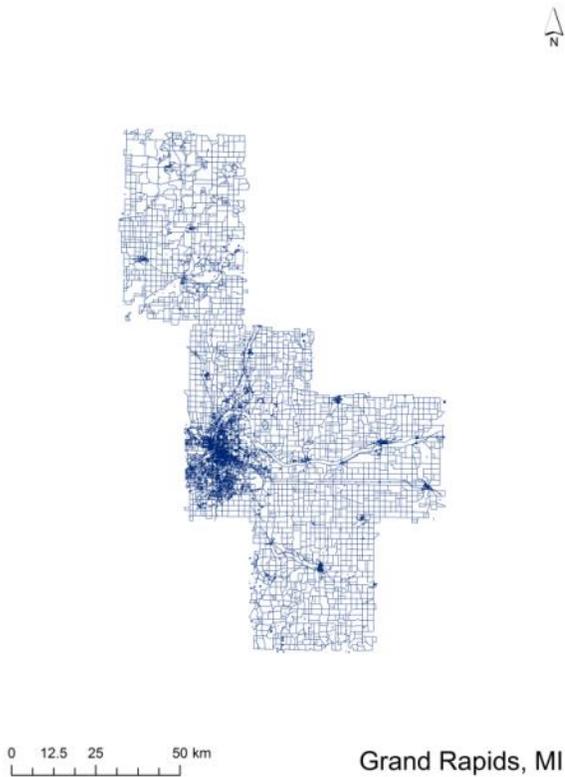

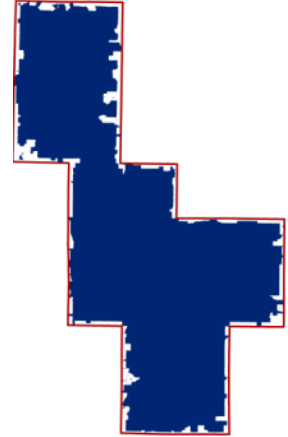

Road Density Map

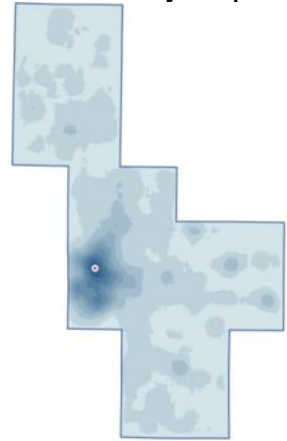

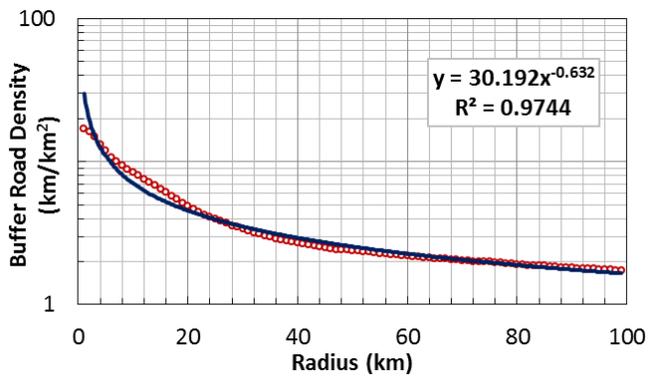

Road Density Fit

### Characteristics

| | |
|---|---:|
| Founded in | 1825 |
| Population | 895227 |
| Pop Density (/km$^2$) | 134.3 |
| Area (km$^2$) | 6665.8 |
| Road Length (km) | 16684.6 |
| # of Intersections | 42990 |
| Area Threshold (m) | 792 |
| Line Threshold (m) | 926 |
| Point Threshold (m) | 395 |
| Density Index (km$^2$) | 30.192 |
| Decay Index (1/km) | 0.632 |

# Hartford, CT

Road Network

Road Polygon Area

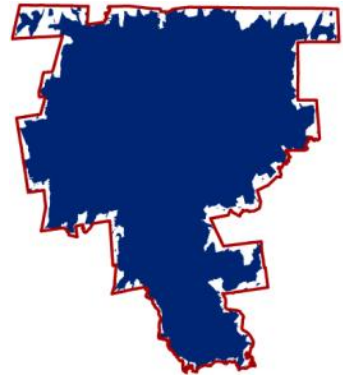

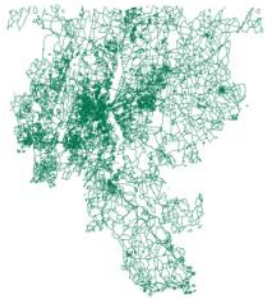

Road Density Map

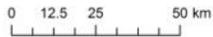

Hartford, CT

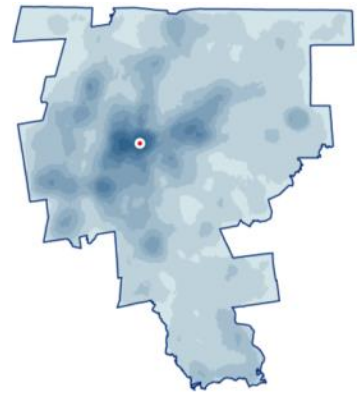

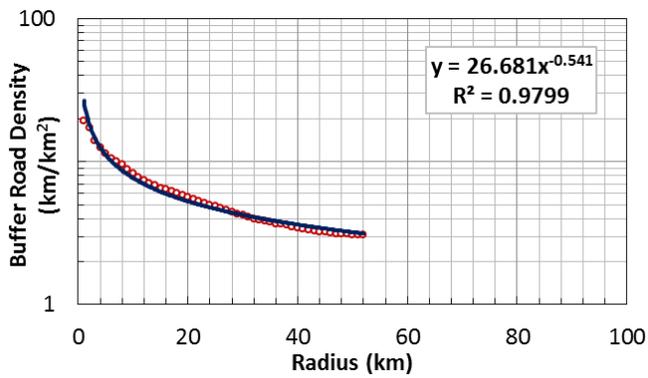

Road Density Fit

### Characteristics

| | |
|---|---:|
| Founded in | 1637 |
| Population | 1400709 |
| Pop Density (/km$^2$) | 401.6 |
| Area (km$^2$) | 3487.6 |
| Road Length (km) | 14992.7 |
| # of Intersections | 56695 |
| Area Threshold (m) | 545 |
| Line Threshold (m) | 514 |
| Point Threshold (m) | 245 |
| Density Index (km$^2$) | 26.681 |
| Decay Index (1/km) | 0.541 |

# Honolulu, HI

Road Network

Road Polygon Area

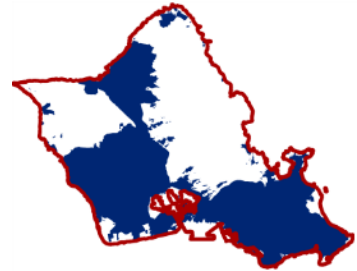

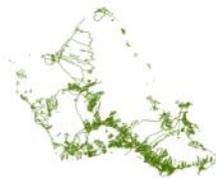

Road Density Map

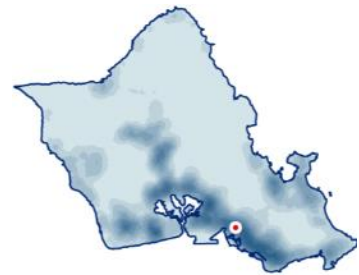

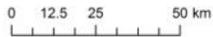

Honolulu, HI

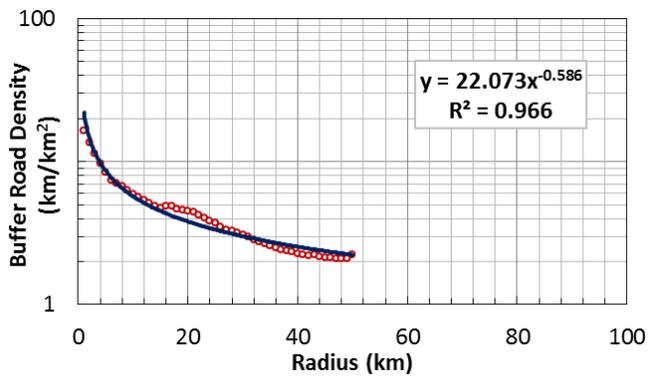

Road Density Fit

### Characteristics

| | |
|---|---:|
| Founded in | 1809 |
| Population | 953207 |
| Pop Density (/km$^2$) | 1229.3 |
| Area (km$^2$) | 775.4 |
| Road Length (km) | 4678.9 |
| # of Intersections | 22904 |
| Area Threshold (m) | 454 |
| Line Threshold (m) | 361 |
| Point Threshold (m) | 178 |
| Density Index (km$^2$) | 22.073 |
| Decay Index (1/km) | 0.586 |

# Houston, TX

Road Network

Road Polygon Area

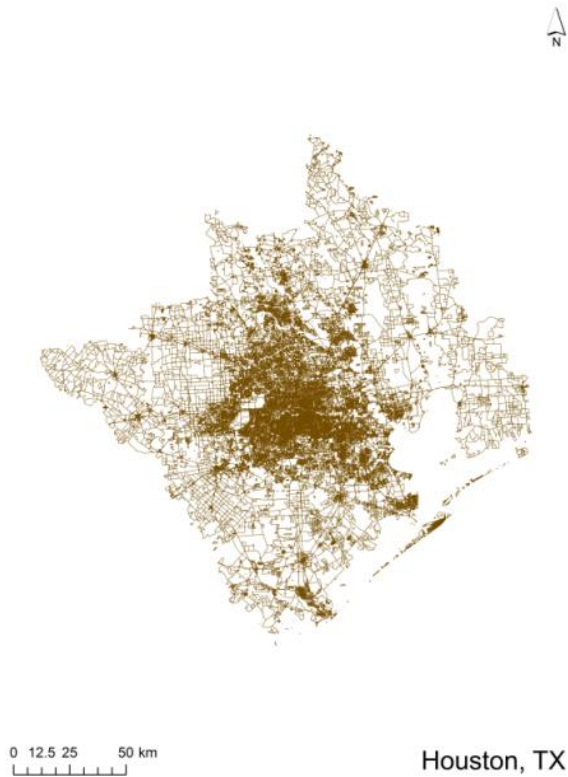

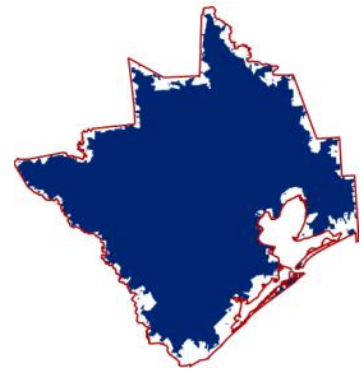

Road Density Map

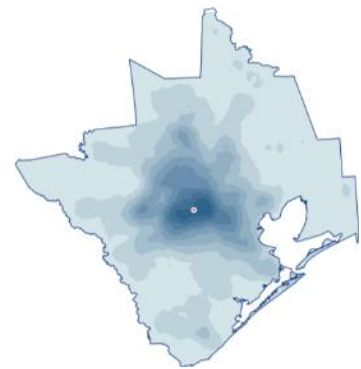

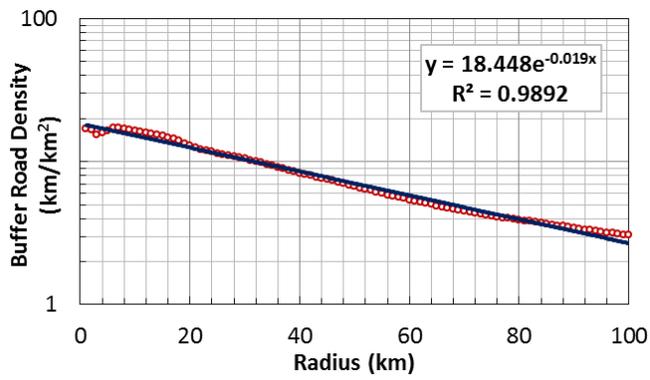

Road Density Fit

### Characteristics

| | |
|---|---:|
| Founded in | 1837 |
| Population | 6052475 |
| Pop Density (/km$^2$) | 294 |
| Area (km$^2$) | 20585.7 |
| Road Length (km) | 83365 |
| # of Intersections | 353831 |
| Area Threshold (m) | 904 |
| Line Threshold (m) | 450 |
| Point Threshold (m) | 210 |
| Density Index (km$^2$) | 18.448 |
| Decay Index (1/km) | 0.019 |

# Indianapolis, IN

Road Network

Road Polygon Area

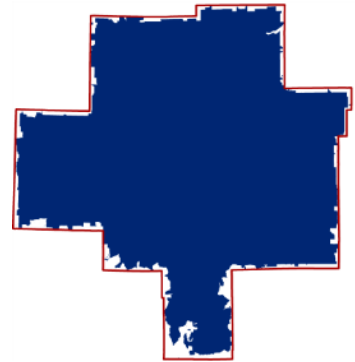

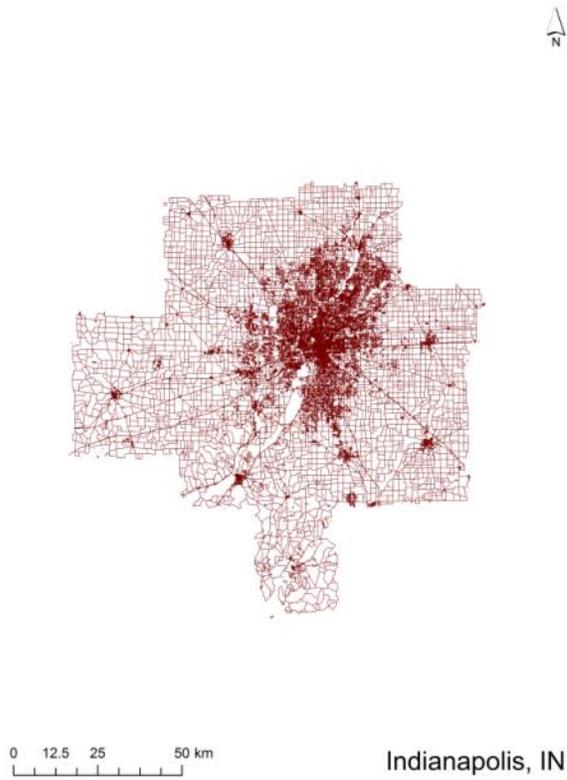

Road Density Map

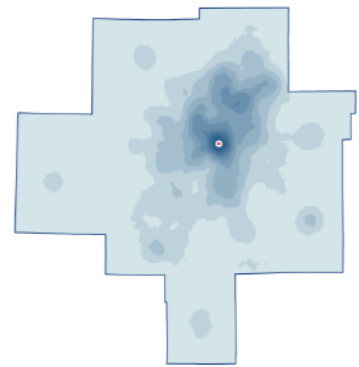

Indianapolis, IN

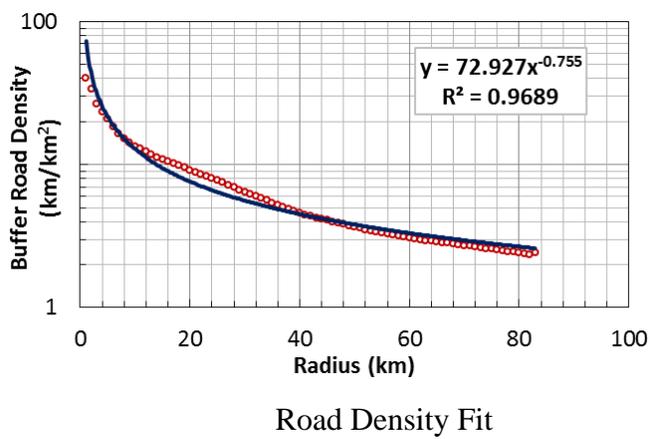

Road Density Fit

### Characteristics

| | |
|---|---:|
| Founded in | 1821 |
| Population | 1856996 |
| Pop Density (/km$^2$) | 199.9 |
| Area (km$^2$) | 9289.1 |
| Road Length (km) | 32389.9 |
| # of Intersections | 150469 |
| Area Threshold (m) | 863 |
| Line Threshold (m) | 575 |
| Point Threshold (m) | 213 |
| Density Index (km$^2$) | 72.927 |
| Decay Index (1/km) | 0.755 |

# Jacksonville, FL

Road Network

Road Polygon Area

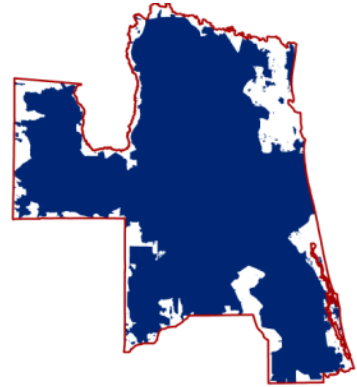

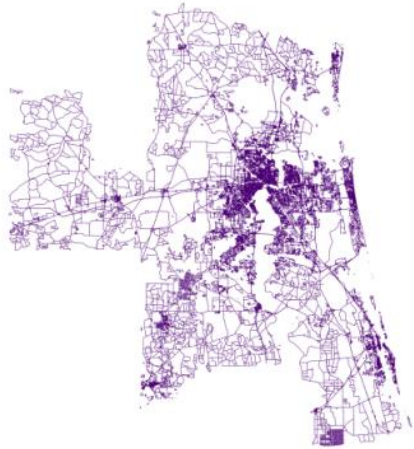

Road Density Map

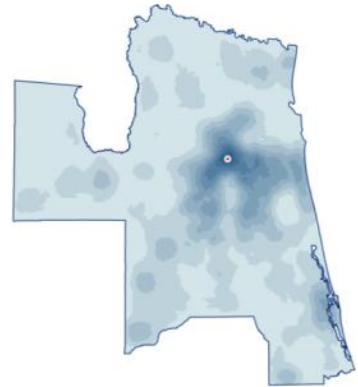

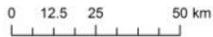

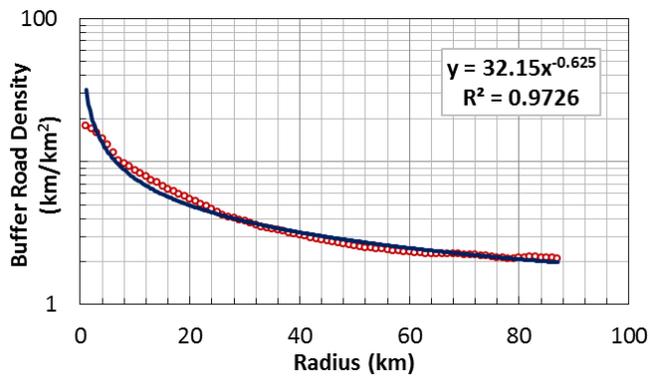

Road Density Fit

## Characteristics

| | |
|---|---:|
| Founded in | 1822 |
| Population | 1451740 |
| Pop Density (/km$^2$) | 202.1 |
| Area (km$^2$) | 7182.3 |
| Road Length (km) | 22067.4 |
| # of Intersections | 76396 |
| Area Threshold (m) | 923 |
| Line Threshold (m) | 670 |
| Point Threshold (m) | 271 |
| Density Index (km$^2$) | 32.15 |
| Decay Index (1/km) | 0.625 |

# Kansas, KS

Road Network

Road Polygon Area

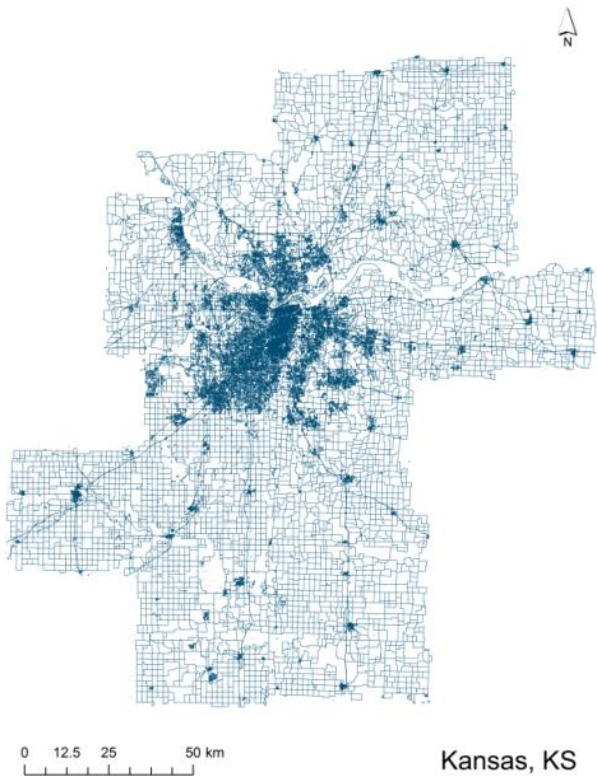

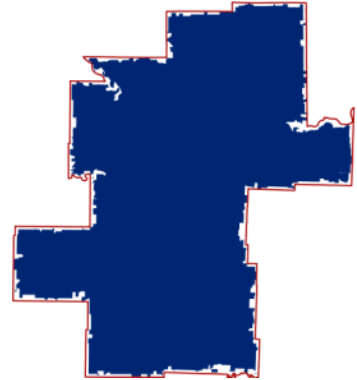

Road Density Map

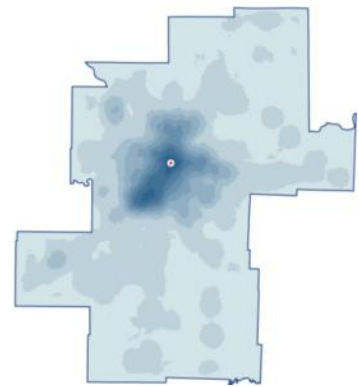

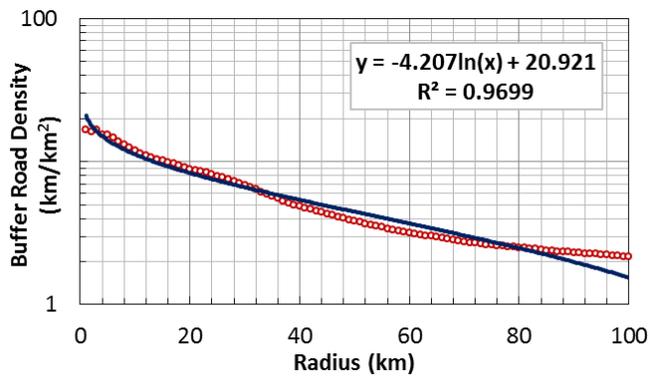

Road Density Fit

### Characteristics

| | |
|---|---:|
| Founded in | 1868 |
| Population | 2138010 |
| Pop Density (/km$^2$) | 111.7 |
| Area (km$^2$) | 19148.1 |
| Road Length (km) | 50639.6 |
| # of Intersections | 184748 |
| Area Threshold (m) | 1028 |
| Line Threshold (m) | 793 |
| Point Threshold (m) | 282 |
| Density Index (km$^2$) | 20.921 |
| Decay Index (1/km) | 4.207 |

# Las Vegas, NV

Road Network

Road Polygon Area

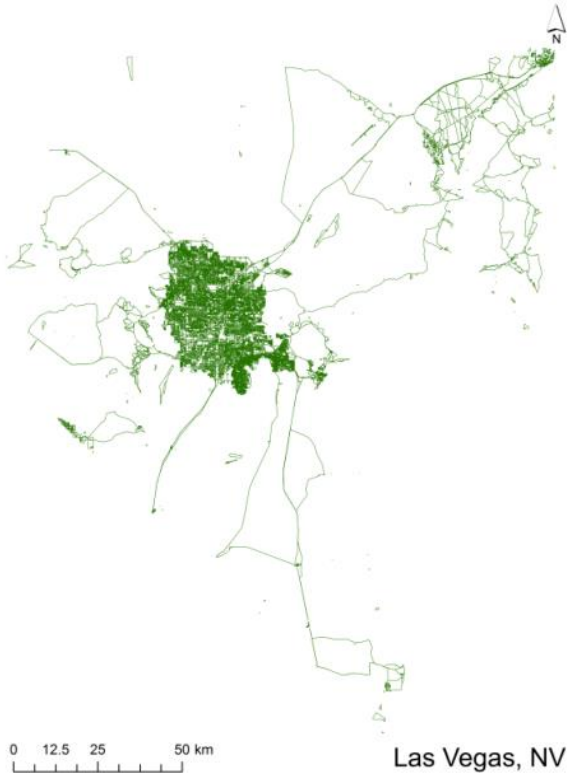

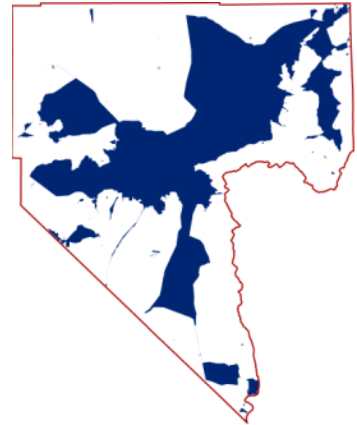

Road Density Map

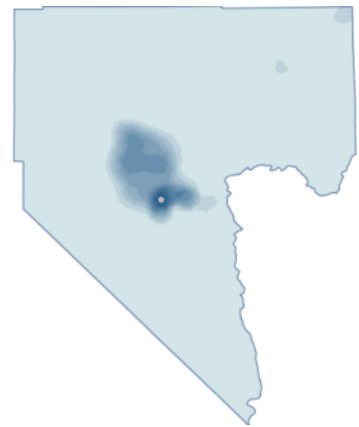

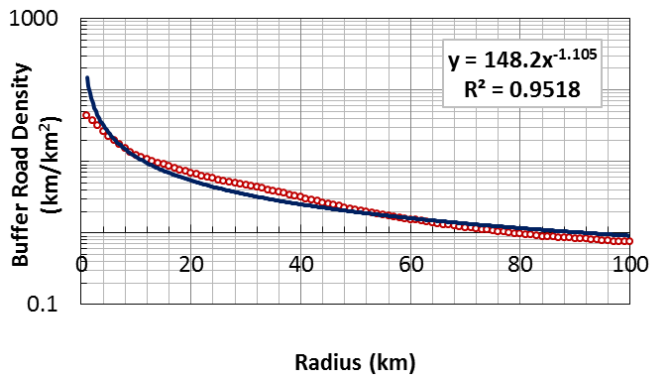

Road Density Fit

### Characteristics

| | |
|---|---:|
| Founded in | 1905 |
| Population | 2010951 |
| Pop Density (/km$^2$) | 274.3 |
| Area (km$^2$) | 7330.1 |
| Road Length (km) | 20926.8 |
| # of Intersections | 104925 |
| Area Threshold (m) | 1330 |
| Line Threshold (m) | 484 |
| Point Threshold (m) | 181 |
| Density Index (km$^2$) | 148.2 |
| Decay Index (1/km) | 1.105 |

# Lewiston, ID

Road Network

Road Polygon Area

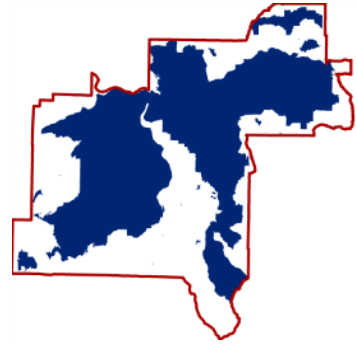

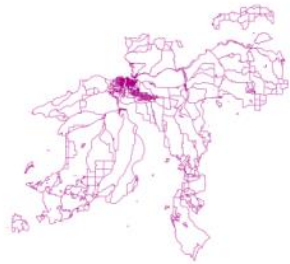

Road Density Map

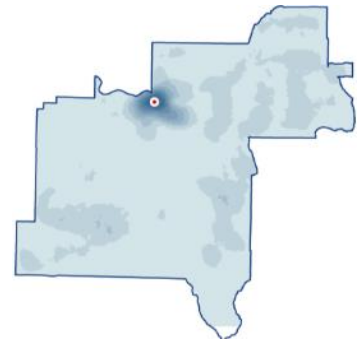

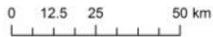

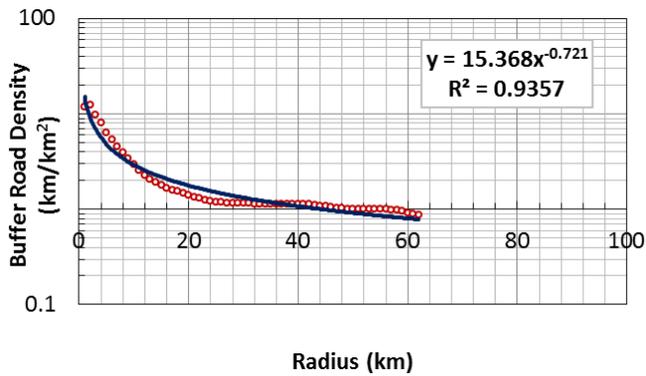

Road Density Fit

### Characteristics

| | |
|---|---:|
| Founded in | 1861 |
| Population | 85096 |
| Pop Density (/km$^2$) | 40.4 |
| Area (km$^2$) | 2104.6 |
| Road Length (km) | 4206.1 |
| # of Intersections | 6334 |
| Area Threshold (m) | 663 |
| Line Threshold (m) | 980 |
| Point Threshold (m) | 727 |
| Density Index (km$^2$) | 15.368 |
| Decay Index (1/km) | 0.721 |

# Los Angeles, CA

Road Network

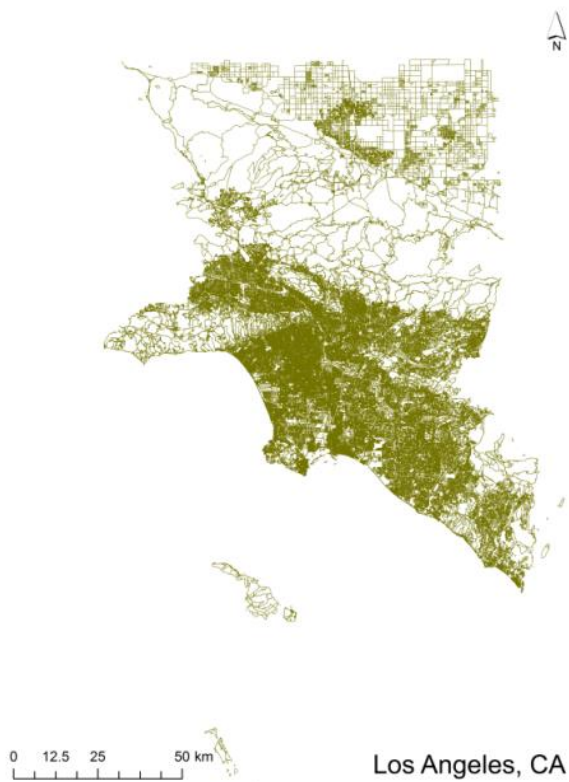

Road Polygon Area

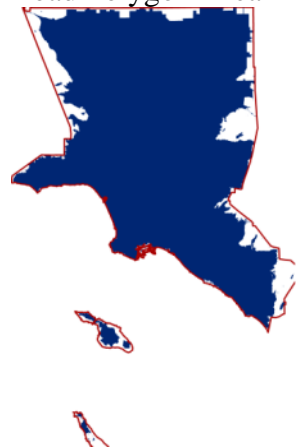

Road Density Map

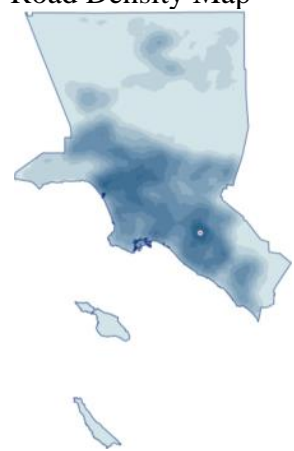

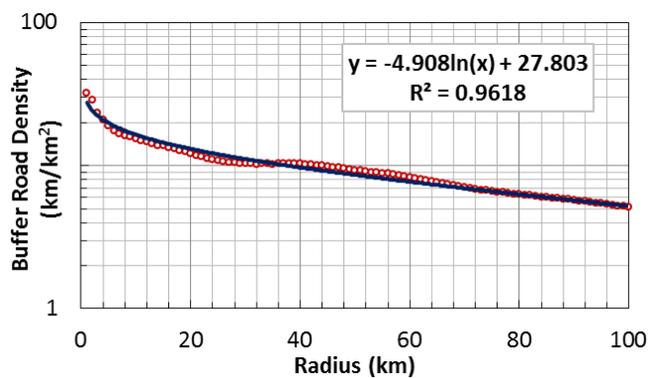

Road Density Fit

### Characteristics

| | |
|---|---:|
| Founded in | 1781 |
| Population | 13059105 |
| Pop Density (/km$^2$) | 1196.6 |
| Area (km$^2$) | 10913.2 |
| Road Length (km) | 70096.7 |
| # of Intersections | 335638 |
| Area Threshold (m) | 962 |
| Line Threshold (m) | 230 |
| Point Threshold (m) | 152 |
| Density Index (km$^2$) | 27.803 |
| Decay Index (1/km) | 4.908 |

# Louisville, KY

Road Network

Road Polygon Area

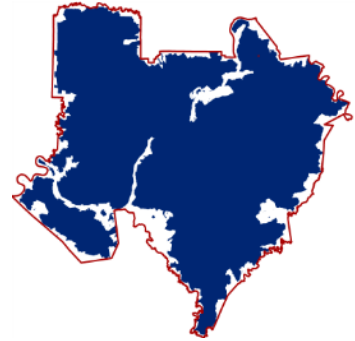

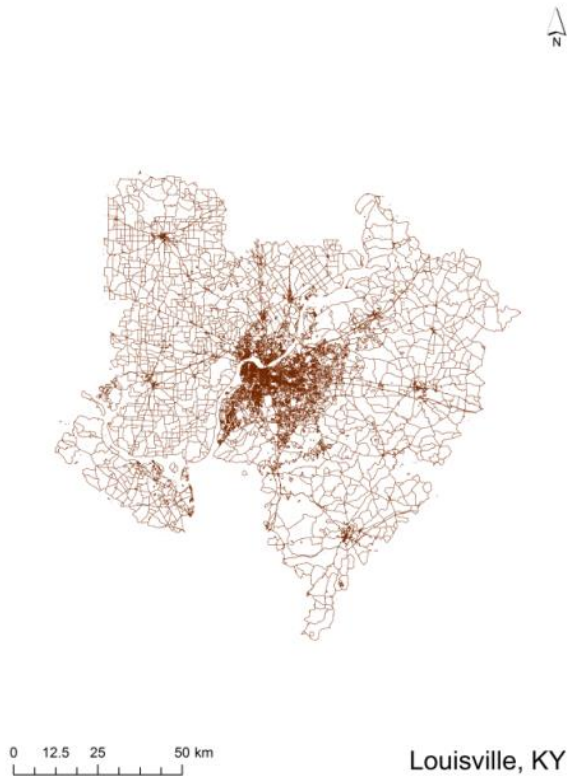

Road Density Map

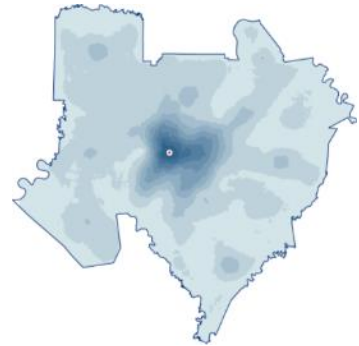

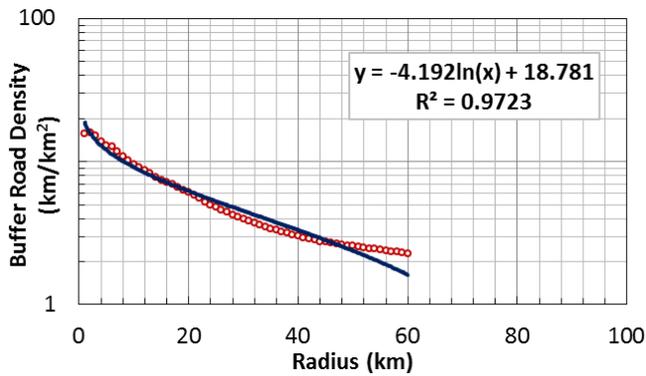

Road Density Fit

$y = -4.192\ln(x) + 18.781$
$R^2 = 0.9723$

## Characteristics

| | |
|---|---:|
| Founded in | 1778 |
| Population | 1443801 |
| Pop Density (/km$^2$) | 156.5 |
| Area (km$^2$) | 9227.8 |
| Road Length (km) | 24453.7 |
| # of Intersections | 82680 |
| Area Threshold (m) | 767 |
| Line Threshold (m) | 879 |
| Point Threshold (m) | 327 |
| Density Index (km$^2$) | 18.781 |
| Decay Index (1/km) | 4.192 |

# Memphis, TN

Road Network

Road Polygon Area

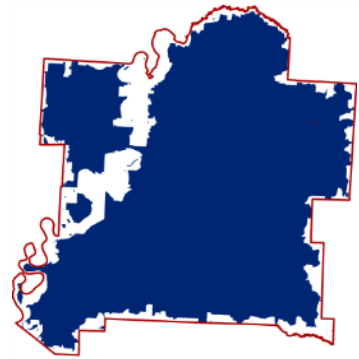

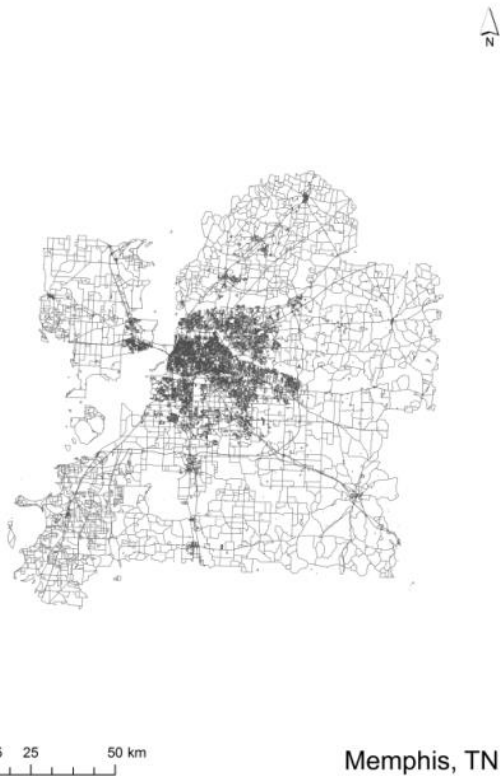

Road Density Map

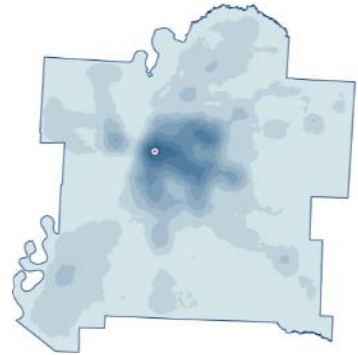

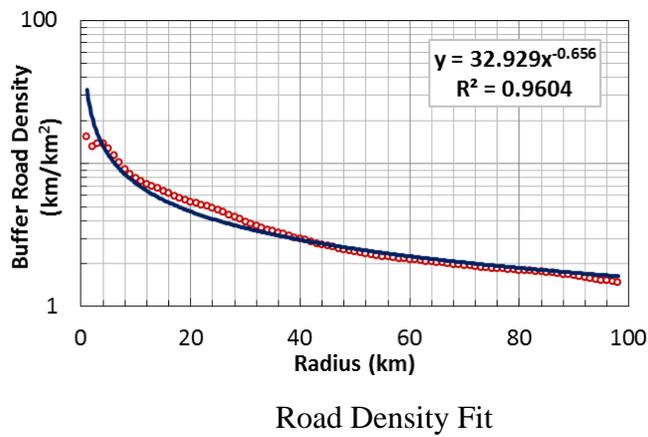

Road Density Fit

### Characteristics

| | |
|---|---:|
| Founded in | 1819 |
| Population | 1398172 |
| Pop Density (/km$^2$) | 139.1 |
| Area (km$^2$) | 10049.2 |
| Road Length (km) | 25028.4 |
| # of Intersections | 74462 |
| Area Threshold (m) | 960 |
| Line Threshold (m) | 891 |
| Point Threshold (m) | 348 |
| Density Index (km$^2$) | 32.929 |
| Decay Index (1/km) | 0.656 |

# Miami, FL

Road Network

Road Polygon Area

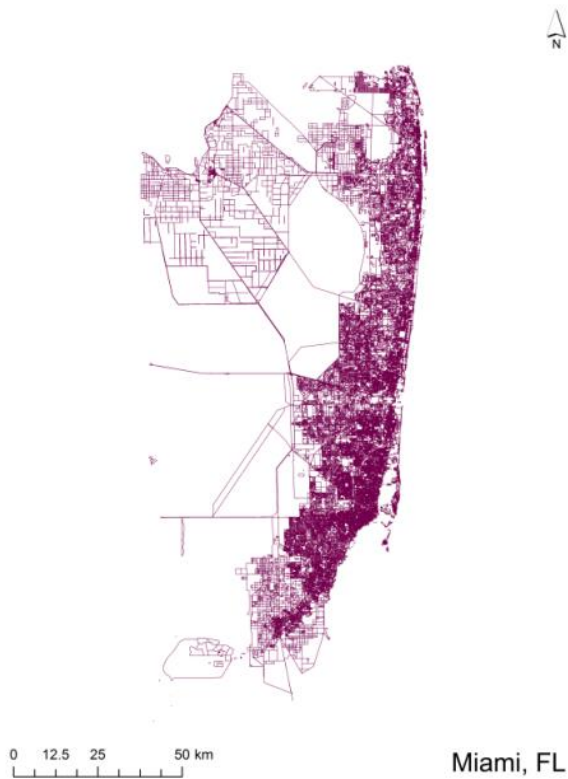

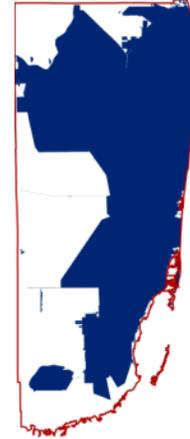

Road Density Map

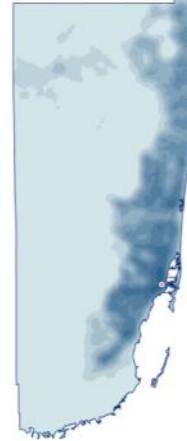

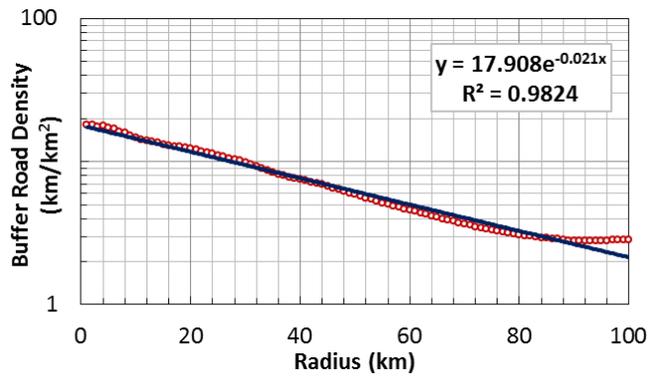

Road Density Fit

## Characteristics

| | |
|---|---:|
| Founded in | 1896 |
| Population | 5571523 |
| Pop Density (/km$^2$) | 662.5 |
| Area (km$^2$) | 8410.3 |
| Road Length (km) | 42827.1 |
| # of Intersections | 178680 |
| Area Threshold (m) | 1660 |
| Line Threshold (m) | 248 |
| Point Threshold (m) | 174 |
| Density Index (km$^2$) | 17.908 |
| Decay Index (1/km) | 0.021 |

# Milwaukee, WI

Road Network

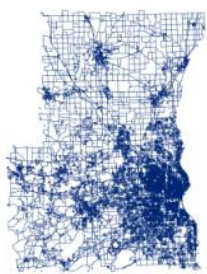

Milwaukee, WI

Road Polygon Area

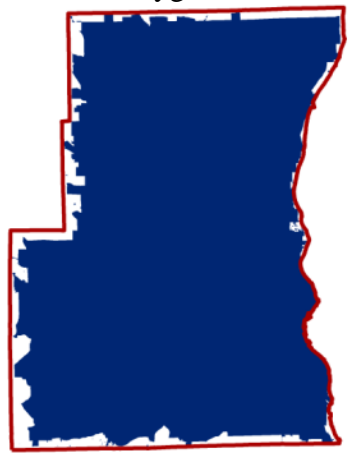

Road Density Map

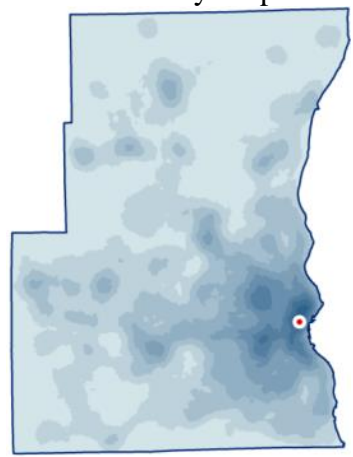

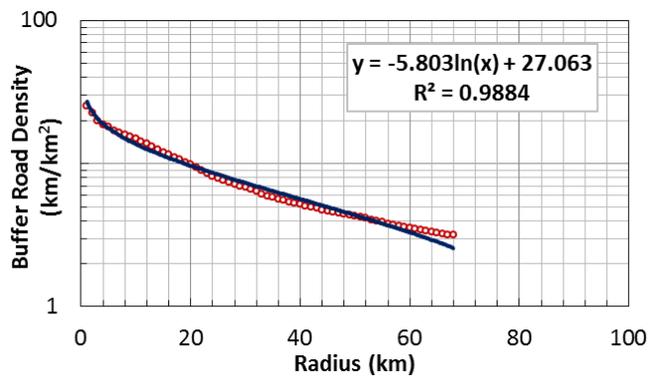

Road Density Fit

## Characteristics

| | |
|---|---:|
| Founded in | 1833 |
| Population | 1602022 |
| Pop Density (/km$^2$) | 456.7 |
| Area (km$^2$) | 3507.8 |
| Road Length (km) | 17207.1 |
| # of Intersections | 66802 |
| Area Threshold (m) | 700 |
| Line Threshold (m) | 386 |
| Point Threshold (m) | 212 |
| Density Index (km$^2$) | 27.063 |
| Decay Index (1/km) | 5.803 |

Fit: $y = -5.803\ln(x) + 27.063$, $R^2 = 0.9884$

# Minneapolis, MN

Road Network

Road Polygon Area

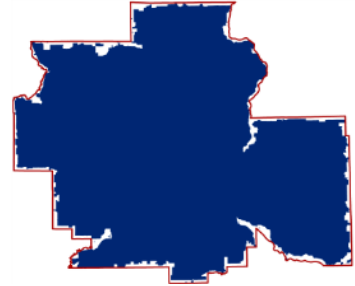

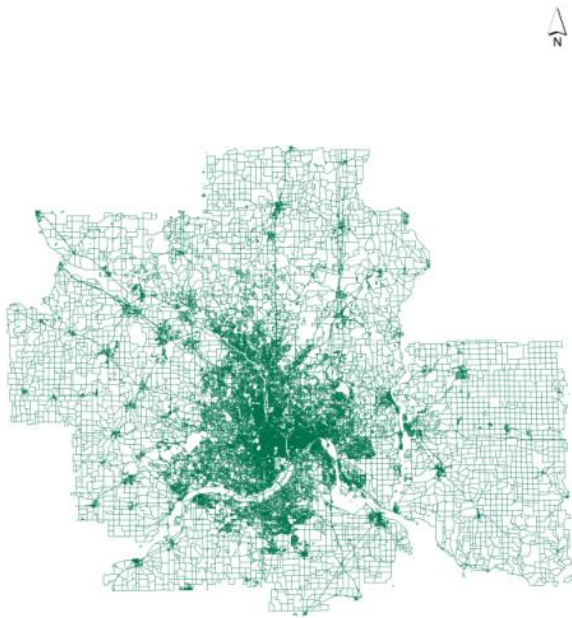

Road Density Map

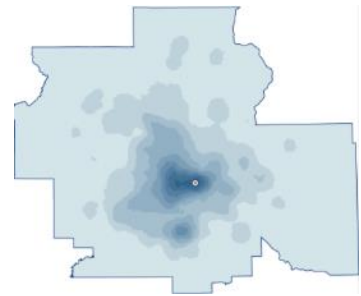

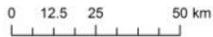
Minneapolis, MN

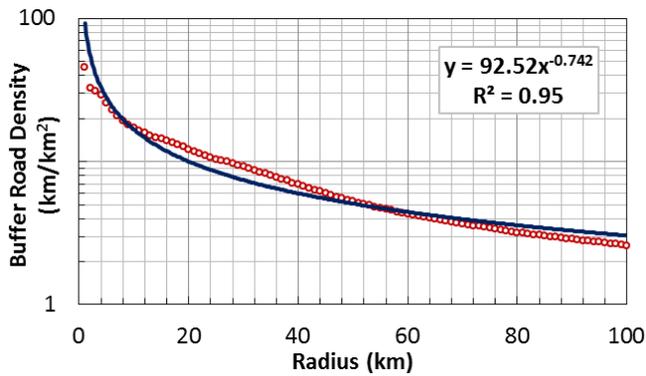

Road Density Fit

### Characteristics

| | |
|---|---:|
| Founded in | 1867 |
| Population | 3412291 |
| Pop Density (/km$^2$) | 222.1 |
| Area (km$^2$) | 15365.8 |
| Road Length (km) | 57532 |
| # of Intersections | 259788 |
| Area Threshold (m) | 904 |
| Line Threshold (m) | 502 |
| Point Threshold (m) | 207 |
| Density Index (km$^2$) | 92.52 |
| Decay Index (1/km) | 0.742 |

# Nashville, TN

Road Network

Road Polygon Area

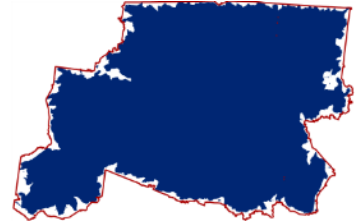

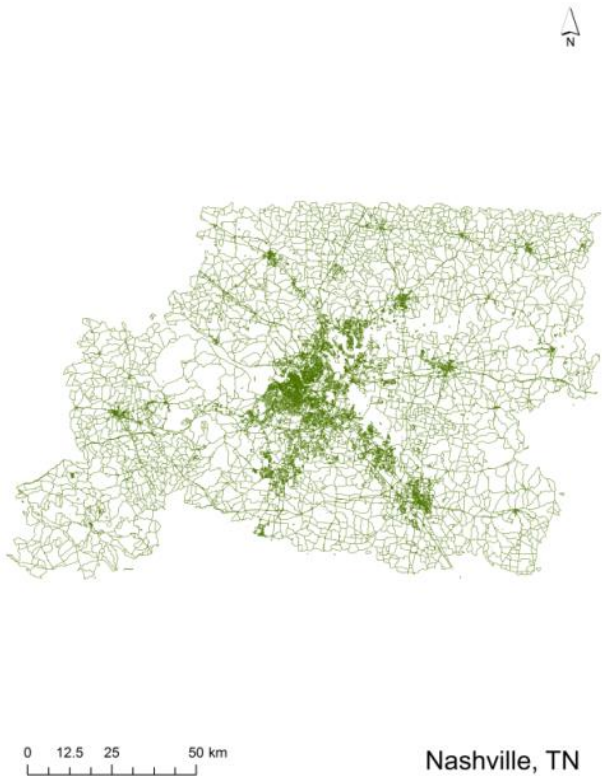

Road Density Map

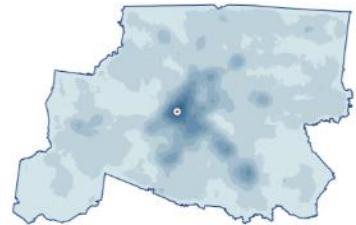

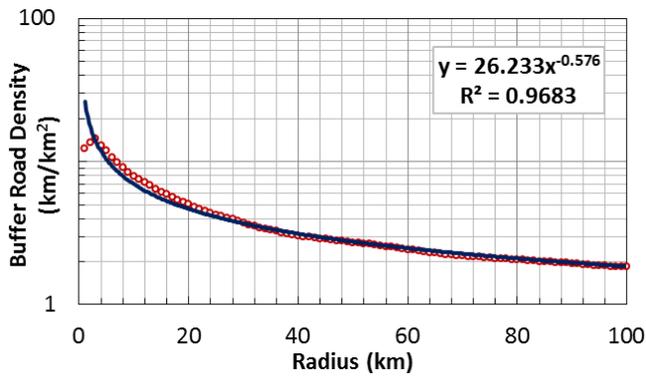

Road Density Fit

### Characteristics

| | |
|---|---:|
| Founded in | 1779 |
| Population | 1740134 |
| Pop Density (/km$^2$) | 128.1 |
| Area (km$^2$) | 13588.3 |
| Road Length (km) | 32653.8 |
| # of Intersections | 90700 |
| Area Threshold (m) | 868 |
| Line Threshold (m) | 919 |
| Point Threshold (m) | 383 |
| Density Index (km$^2$) | 26.233 |
| Decay Index (1/km) | 0.576 |

# **New Orleans, LA**

Road Network

Road Polygon Area

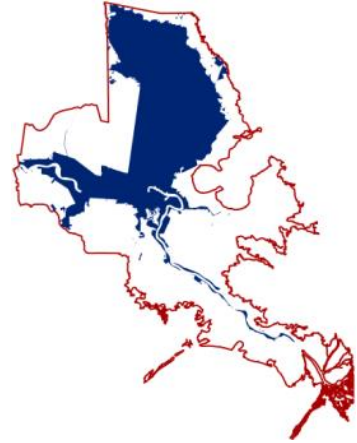

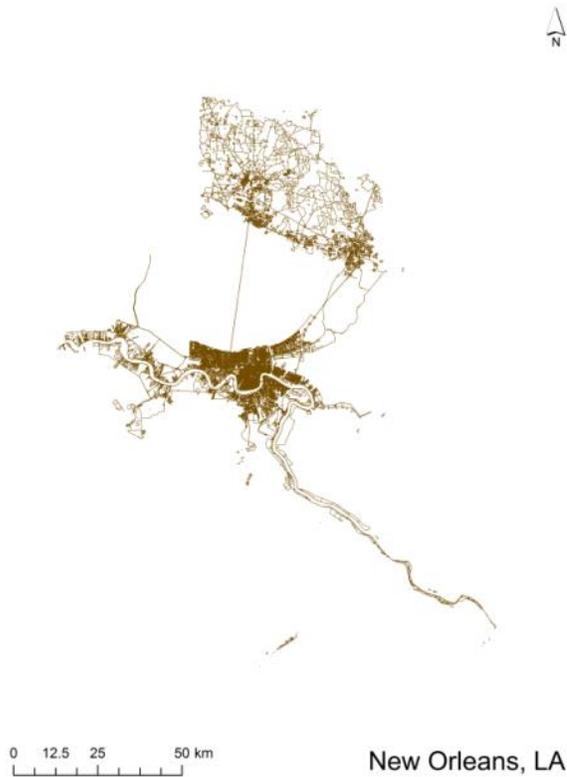

Road Density Map

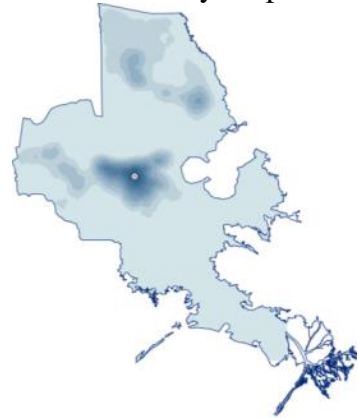

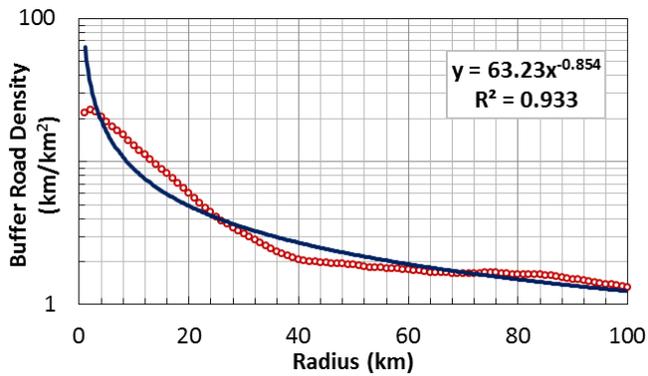

Road Density Fit

### **Characteristics**

| | |
|---|---:|
| Founded in | 1718 |
| Population | 1247062 |
| Pop Density (/km$^2$) | 335.6 |
| Area (km$^2$) | 3715.5 |
| Road Length (km) | 18340.7 |
| # of Intersections | 83361 |
| Area Threshold (m) | 699 |
| Line Threshold (m) | 372 |
| Point Threshold (m) | 189 |
| Density Index (km$^2$) | 63.23 |
| Decay Index (1/km) | 0.854 |

# New York, NY

Road Network

Road Polygon Area

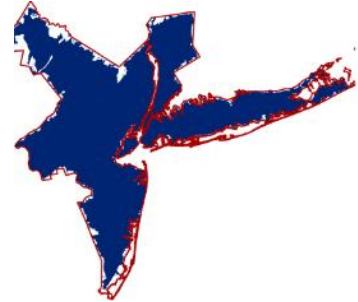

Road Density Map

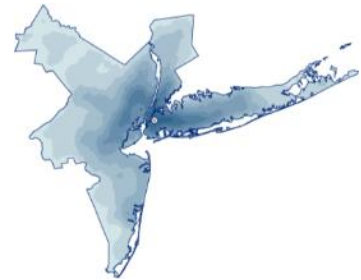

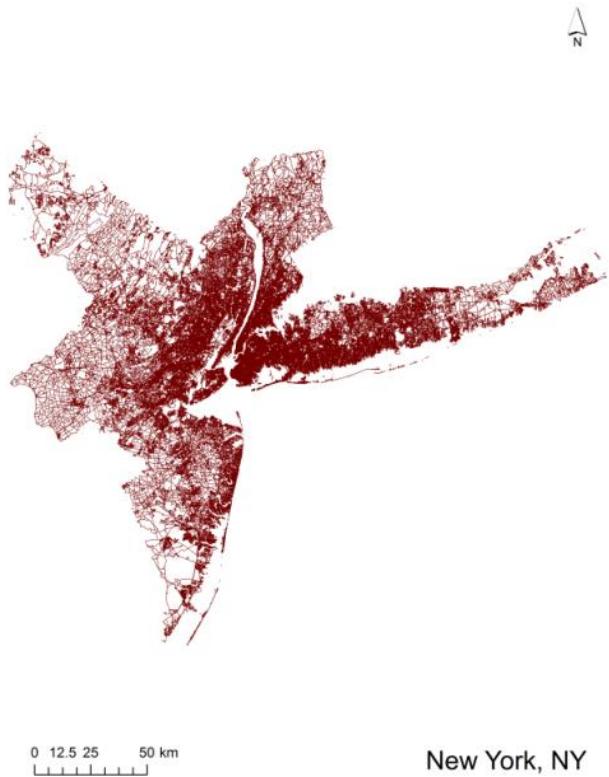

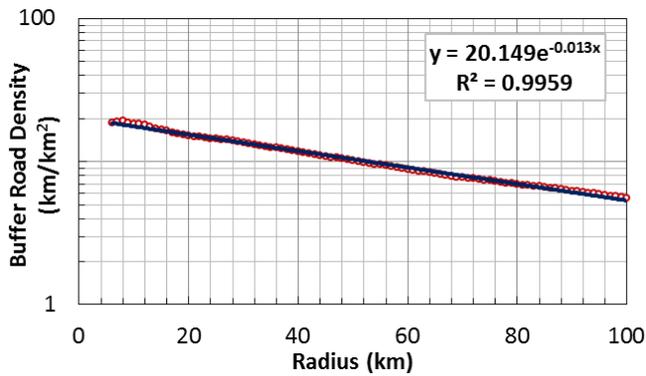

Road Density Fit

### Characteristics

| | |
|---|---:|
| Founded in | 1624 |
| Population | 19217139 |
| Pop Density (/km$^2$) | 1235.7 |
| Area (km$^2$) | 15551.5 |
| Road Length (km) | 105344 |
| # of Intersections | 499969 |
| Area Threshold (m) | 501 |
| Line Threshold (m) | 282 |
| Point Threshold (m) | 170 |
| Density Index (km$^2$) | 20.149 |
| Decay Index (1/km) | 0.013 |

# Oklahoma, OK

Road Network

Road Polygon Area

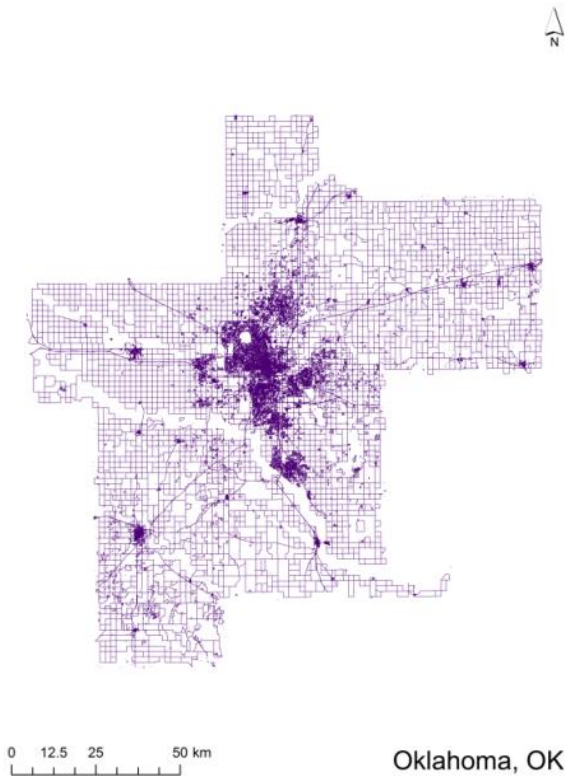

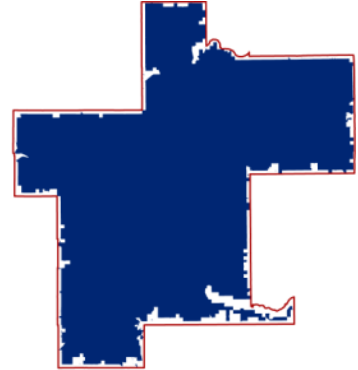

Road Density Map

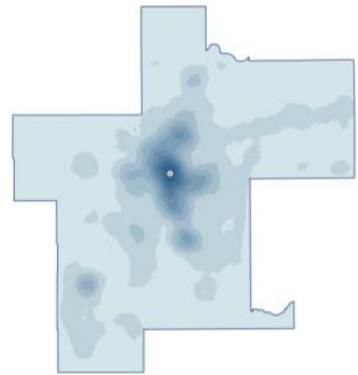

Oklahoma, OK

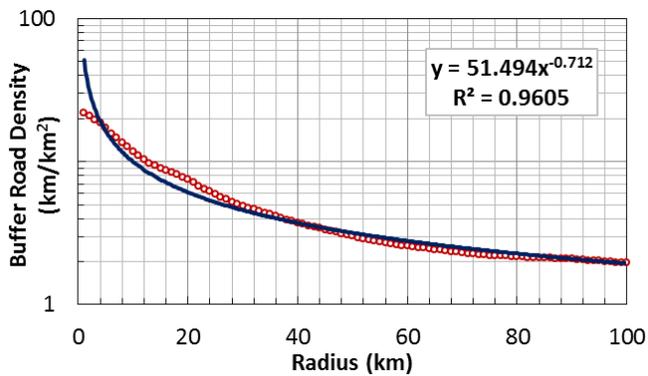

Road Density Fit

### Characteristics

| | |
|---|---:|
| Founded in | 1889 |
| Population | 1359027 |
| Pop Density (/km$^2$) | 104.1 |
| Area (km$^2$) | 13051 |
| Road Length (km) | 34167.6 |
| # of Intersections | 120303 |
| Area Threshold (m) | 955 |
| Line Threshold (m) | 828 |
| Point Threshold (m) | 296 |
| Density Index (km$^2$) | 51.494 |
| Decay Index (1/km) | 0.712 |

# **Orlando, FL**

Road Network

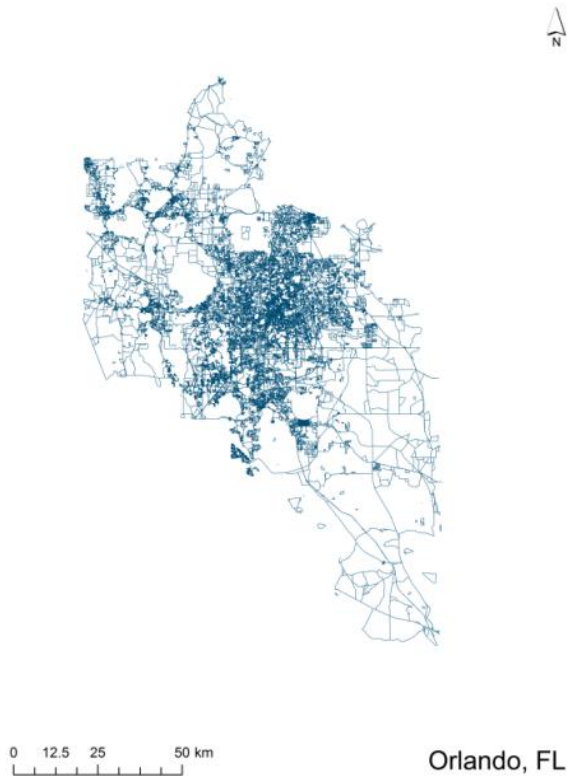

Orlando, FL

Road Polygon Area

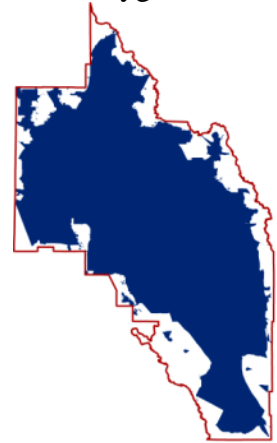

Road Density Map

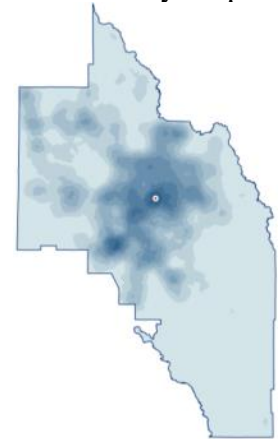

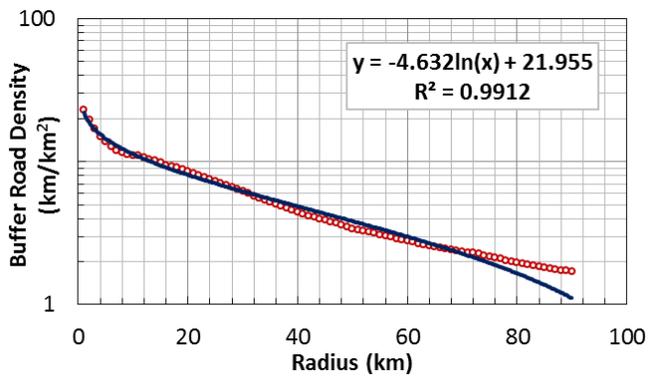

Road Density Fit

### **Characteristics**

| | |
|---|---:|
| Founded in | 1875 |
| Population | 2257901 |
| Pop Density (/km$^2$) | 282.4 |
| Area (km$^2$) | 7996.6 |
| Road Length (km) | 28876.5 |
| # of Intersections | 123076 |
| Area Threshold (m) | 1374 |
| Line Threshold (m) | 418 |
| Point Threshold (m) | 200 |
| Density Index (km$^2$) | 21.955 |
| Decay Index (1/km) | 4.632 |

# Philadelphia, PA

Road Network

Road Polygon Area

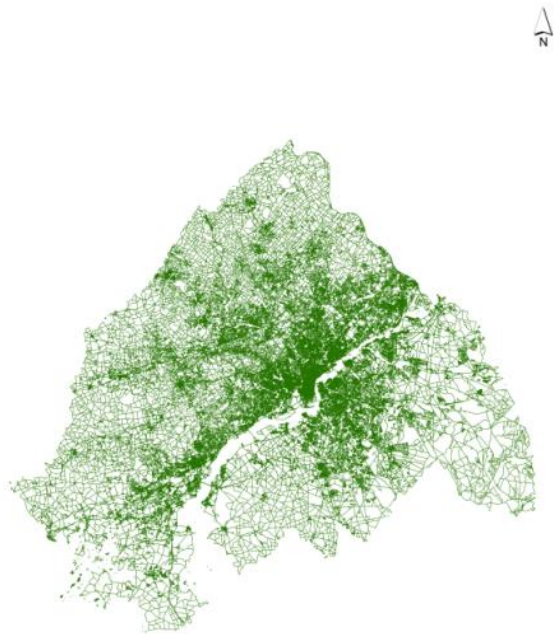

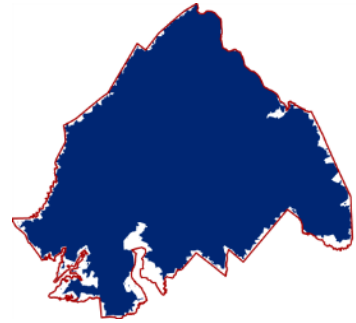

Road Density Map

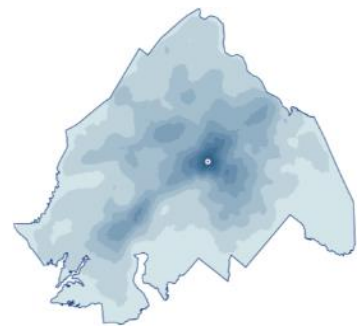

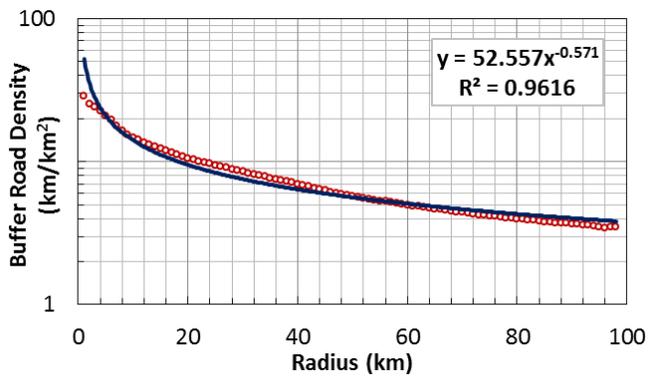

Road Density Fit

### Characteristics

| | |
|---|---:|
| Founded in | 1682 |
| Population | 6234336 |
| Pop Density (/km$^2$) | 553.1 |
| Area (km$^2$) | 11271.7 |
| Road Length (km) | 58104.3 |
| # of Intersections | 256023 |
| Area Threshold (m) | 648 |
| Line Threshold (m) | 378 |
| Point Threshold (m) | 197 |
| Density Index (km$^2$) | 52.557 |
| Decay Index (1/km) | 0.571 |

# Phoenix, AZ

Road Network

Road Polygon Area

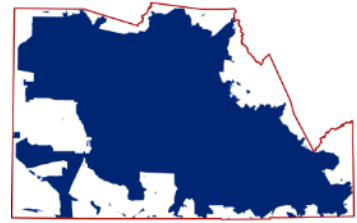

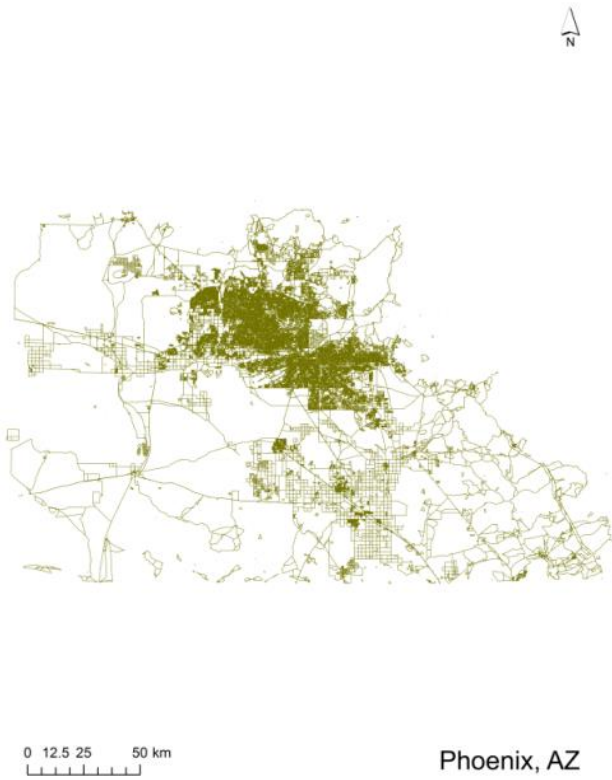

Road Density Map

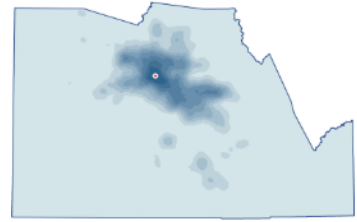

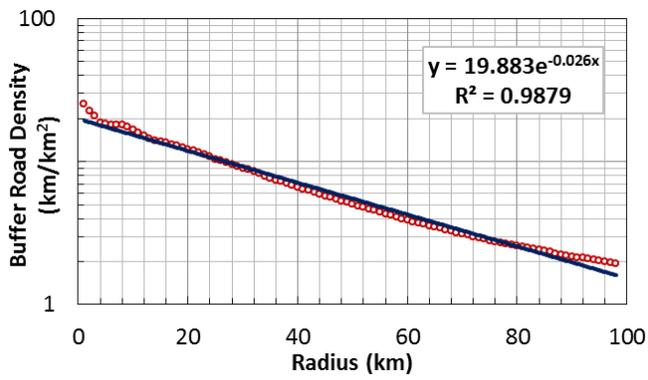

Road Density Fit

### Characteristics

| | |
|---|---:|
| Founded in | 1868 |
| Population | 4262838 |
| Pop Density (/km$^2$) | 165.5 |
| Area (km$^2$) | 25763 |
| Road Length (km) | 60738.6 |
| # of Intersections | 241836 |
| Area Threshold (m) | 1200 |
| Line Threshold (m) | 535 |
| Point Threshold (m) | 221 |
| Density Index (km$^2$) | 19.883 |
| Decay Index (1/km) | 0.026 |

# **Pittsburgh, PA**

Road Network

Road Polygon Area

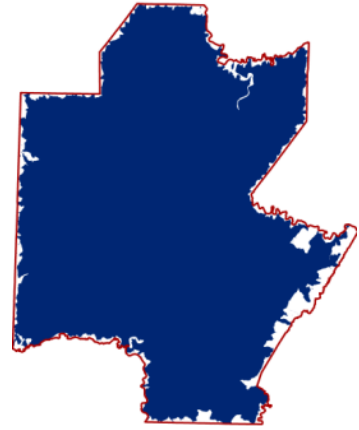

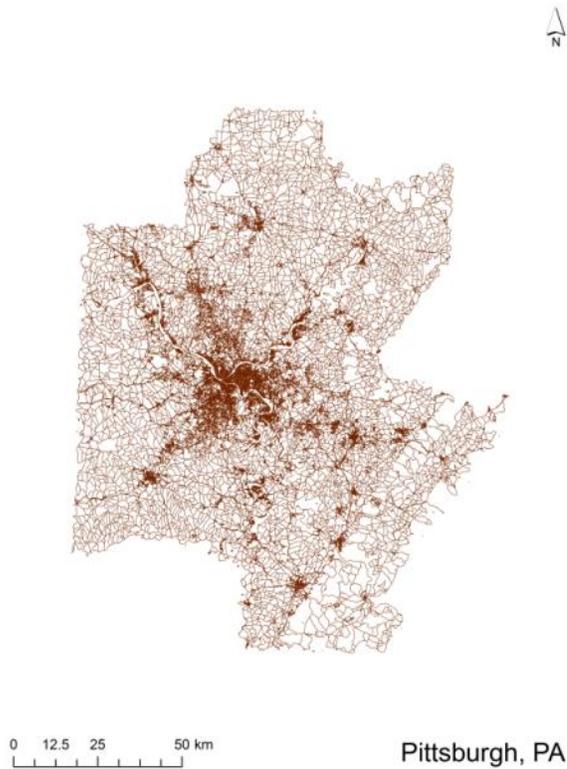

Road Density Map

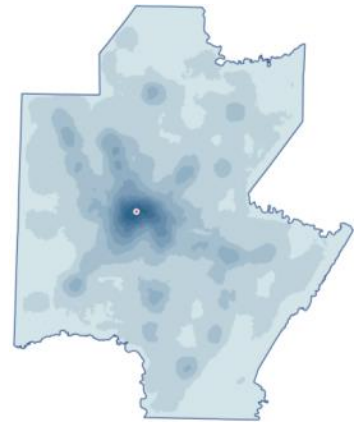

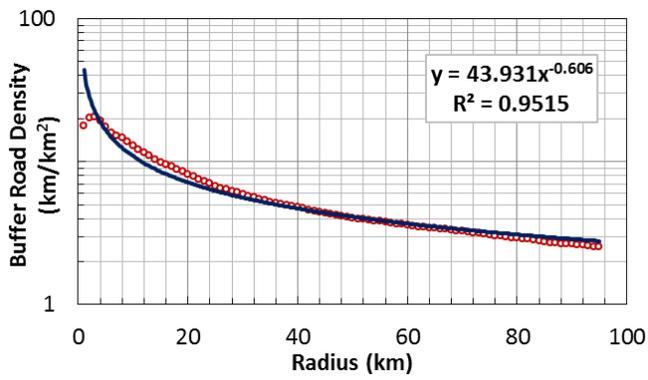

Road Density Fit

### **Characteristics**

| | |
|---|---:|
| Founded in | 1758 |
| Population | 2503836 |
| Pop Density (/km$^2$) | 194.7 |
| Area (km$^2$) | 12859.9 |
| Road Length (km) | 45196.4 |
| # of Intersections | 167027 |
| Area Threshold (m) | 707 |
| Line Threshold (m) | 596 |
| Point Threshold (m) | 267 |
| Density Index (km$^2$) | 43.931 |
| Decay Index (1/km) | 0.606 |

Fit equation: $y = 43.931 x^{-0.606}$, $R^2 = 0.9515$

# Portland, OR

Road Network

Road Polygon Area

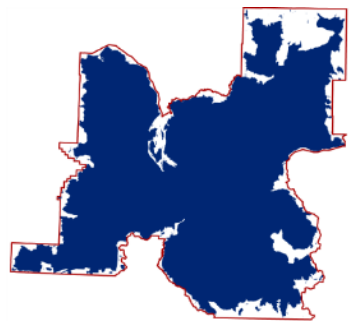

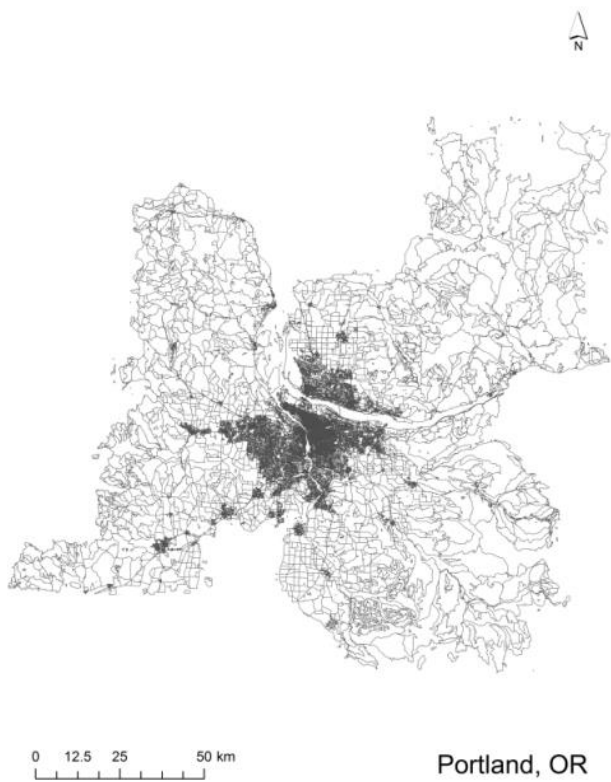

Road Density Map

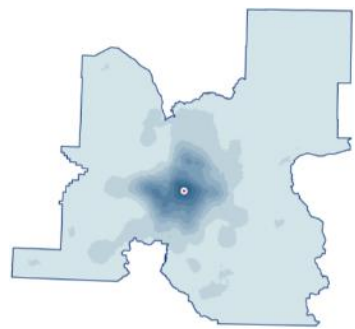

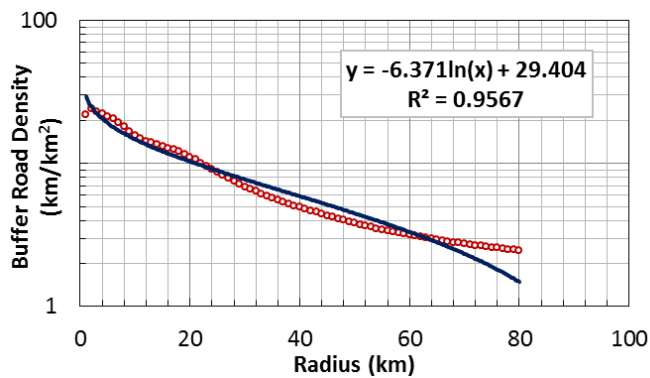

Road Density Fit

## Characteristics

| | |
|---|---:|
| Founded in | 1845 |
| Population | 2363554 |
| Pop Density (/km$^2$) | 161.1 |
| Area (km$^2$) | 14669.4 |
| Road Length (km) | 44544 |
| # of Intersections | 174765 |
| Area Threshold (m) | 787 |
| Line Threshold (m) | 722 |
| Point Threshold (m) | 270 |
| Density Index (km$^2$) | 29.404 |
| Decay Index (1/km) | 6.371 |

# **Providence, RI**

Road Network

Road Polygon Area

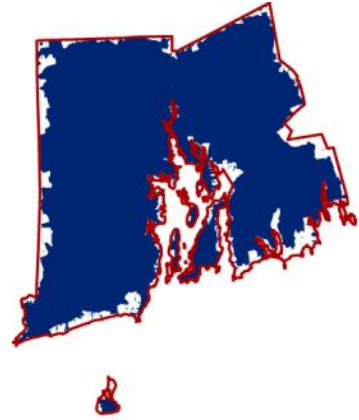

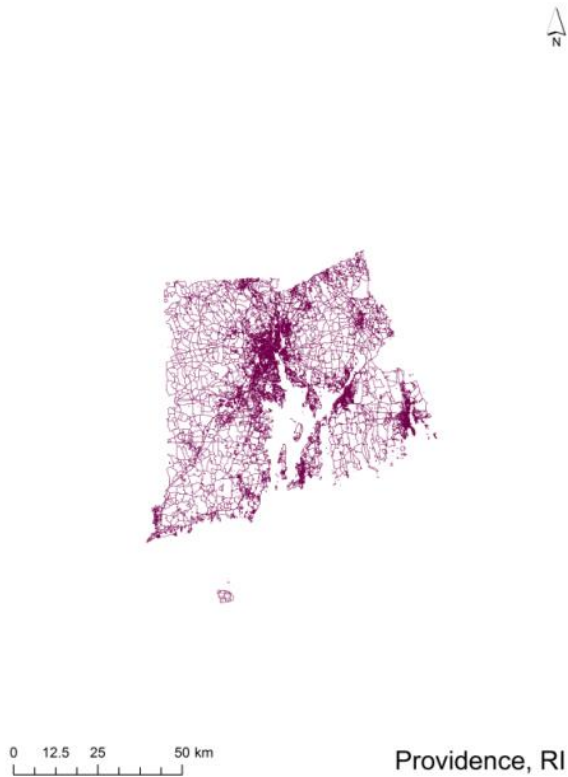

Road Density Map

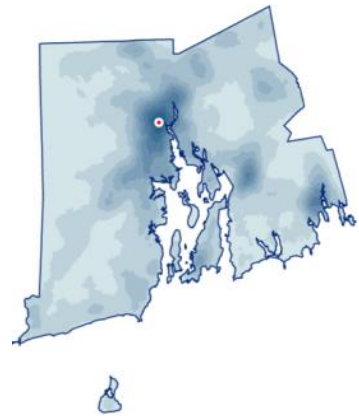

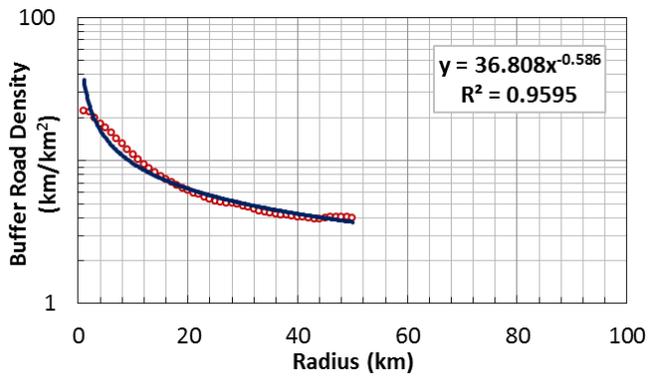

Road Density Fit

### **Characteristics**

| | |
|---|---:|
| Founded in | 1636 |
| Population | 1695760 |
| Pop Density (/km$^2$) | 449.4 |
| Area (km$^2$) | 3773.5 |
| Road Length (km) | 18431.5 |
| # of Intersections | 83871 |
| Area Threshold (m) | 531 |
| Line Threshold (m) | 444 |
| Point Threshold (m) | 201 |
| Density Index (km$^2$) | 36.808 |
| Decay Index (1/km) | 0.586 |

# **Raleigh, NC**

Road Network

Road Polygon Area

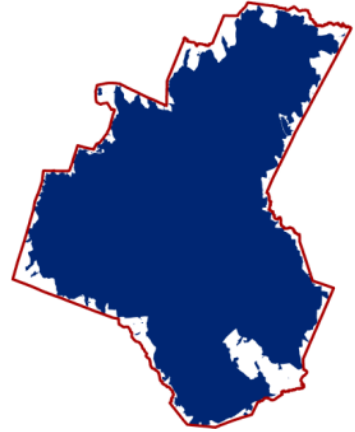

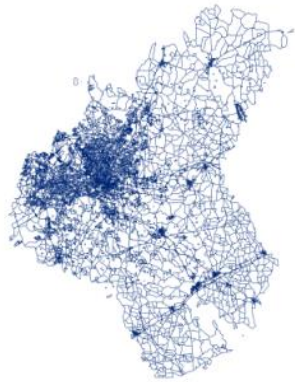

Road Density Map

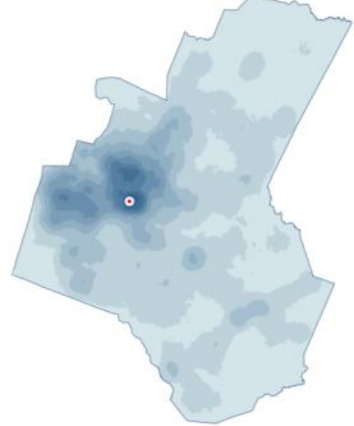

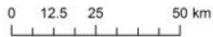

Raliegh, NC

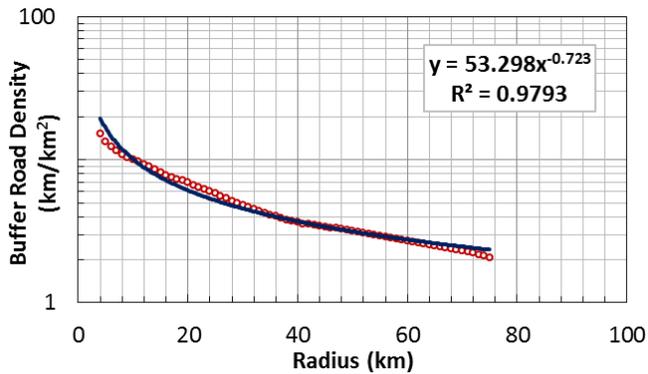

Road Density Fit

### **Characteristics**

| | |
|---|---:|
| Founded in | 1792 |
| Population | 1258825 |
| Pop Density (/km$^2$) | 260.6 |
| Area (km$^2$) | 4830.5 |
| Road Length (km) | 18678 |
| # of Intersections | 81802 |
| Area Threshold (m) | 637 |
| Line Threshold (m) | 562 |
| Point Threshold (m) | 231 |
| Density Index (km$^2$) | 53.298 |
| Decay Index (1/km) | 0.723 |

# Rochester, NY

Road Network

Road Polygon Area

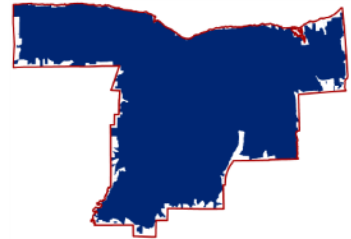

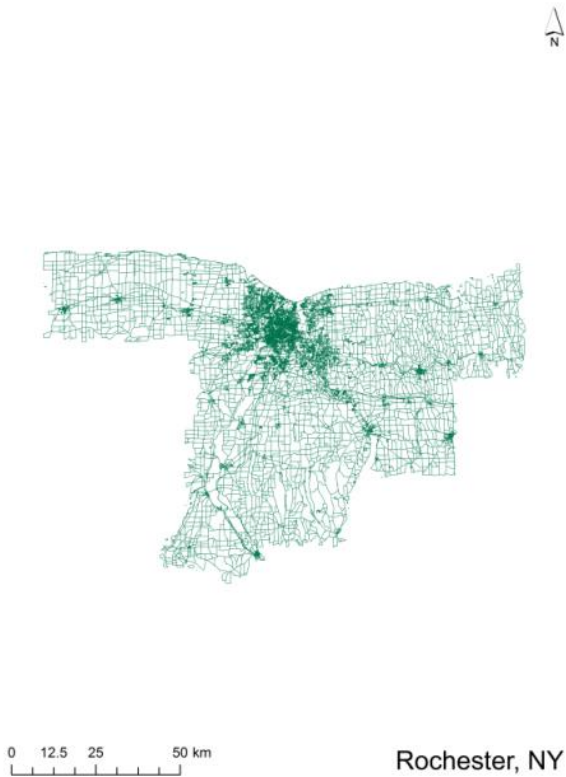

Road Density Map

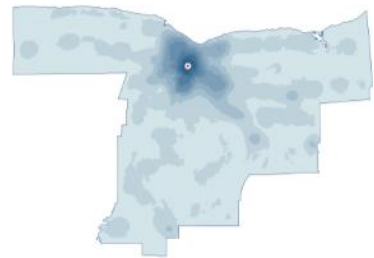

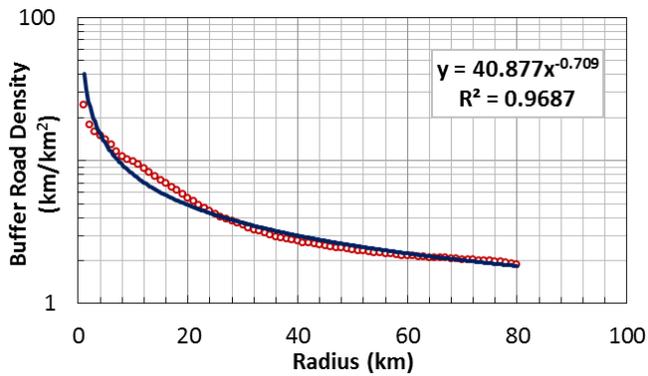

Road Density Fit

**Characteristics**

| | |
|---|---:|
| Founded in | 1803 |
| Population | 1159166 |
| Pop Density (/km$^2$) | 164.7 |
| Area (km$^2$) | 7037.2 |
| Road Length (km) | 17863.9 |
| # of Intersections | 47275 |
| Area Threshold (m) | 881 |
| Line Threshold (m) | 874 |
| Point Threshold (m) | 380 |
| Density Index (km$^2$) | 40.877 |
| Decay Index (1/km) | 0.709 |

# Sacramento, CA

Road Network

Road Polygon Area

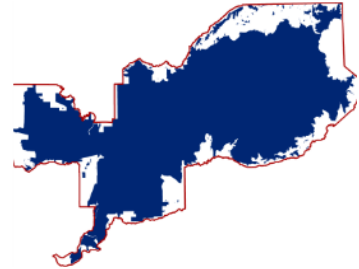

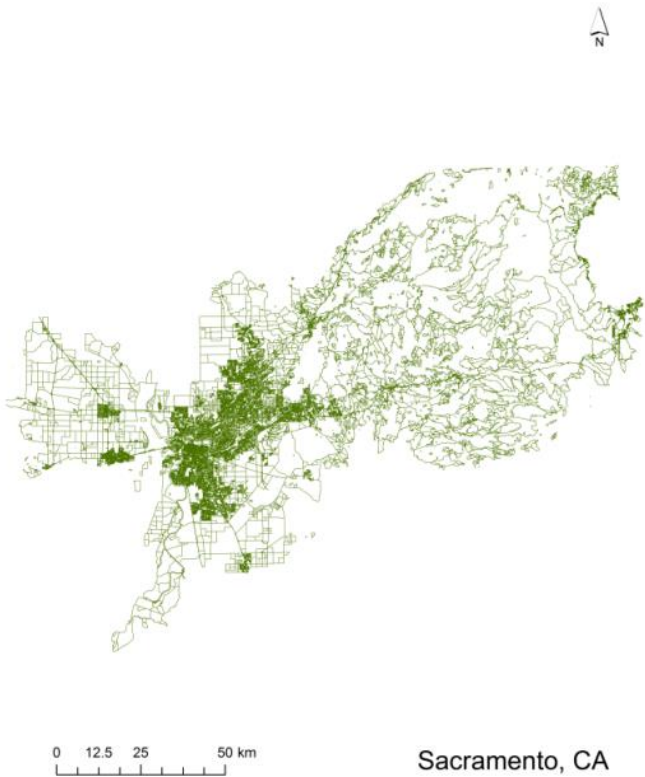

Road Density Map

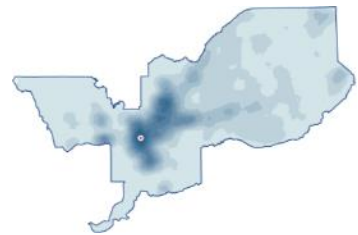

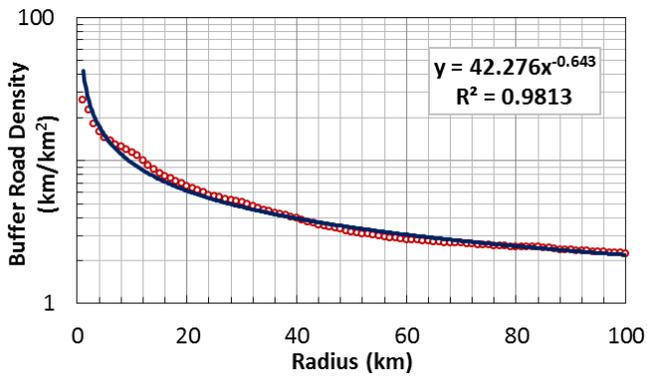

Road Density Fit

### Characteristics

| | |
|---|---:|
| Founded in | 1839 |
| Population | 2277843 |
| Pop Density (/km$^2$) | 224 |
| Area (km$^2$) | 10167 |
| Road Length (km) | 34020.6 |
| # of Intersections | 124839 |
| Area Threshold (m) | 821 |
| Line Threshold (m) | 627 |
| Point Threshold (m) | 260 |
| Density Index (km$^2$) | 42.276 |
| Decay Index (1/km) | 0.643 |

# Salt Lake, UT

Road Network

Road Polygon Area

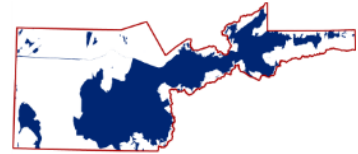

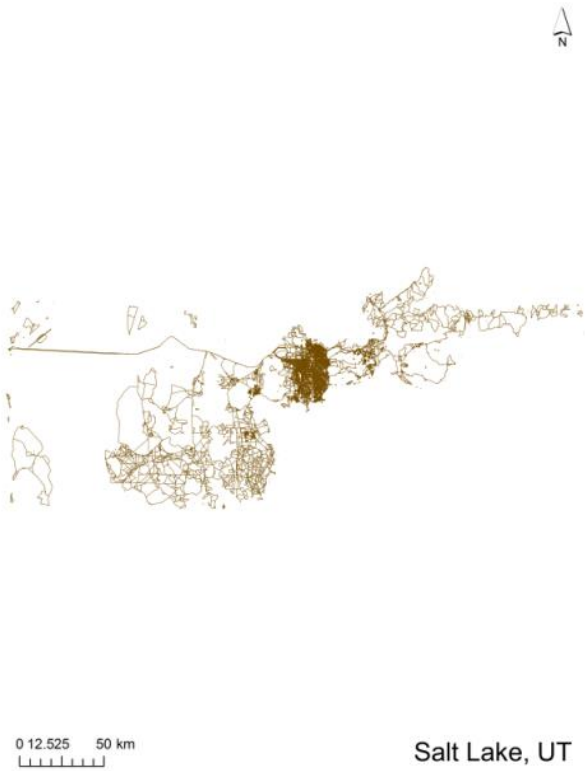

Road Density Map

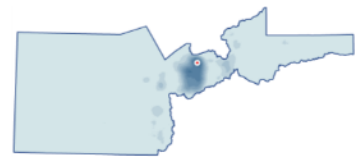

Salt Lake, UT

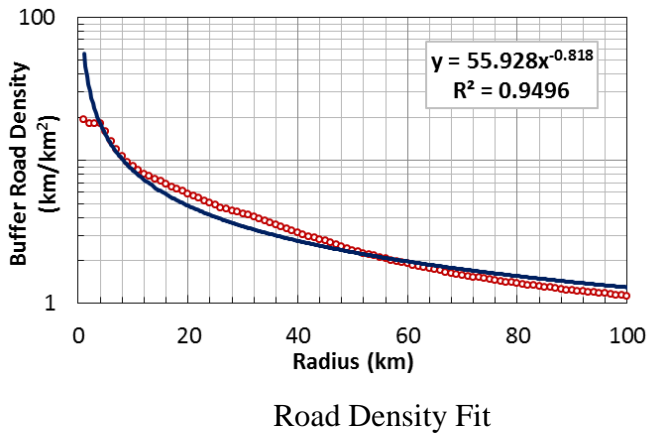

Road Density Fit

### Characteristics

| | |
|---|---:|
| Founded in | 1847 |
| Population | 1246208 |
| Pop Density (/km$^2$) | 114.4 |
| Area (km$^2$) | 10895.1 |
| Road Length (km) | 22387 |
| # of Intersections | 59736 |
| Area Threshold (m) | 1033 |
| Line Threshold (m) | 957 |
| Point Threshold (m) | 397 |
| Density Index (km$^2$) | 55.928 |
| Decay Index (1/km) | 0.818 |

# San Antonio, TX

Road Network

Road Polygon Area

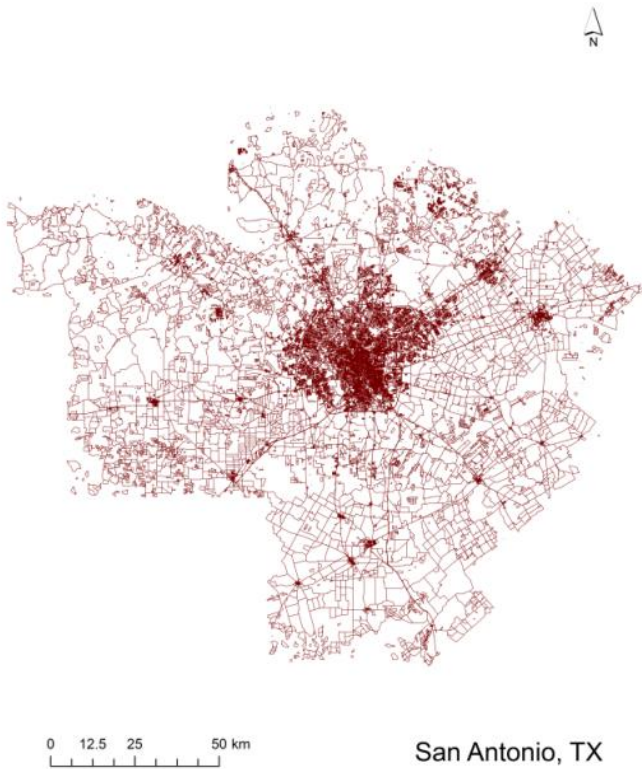

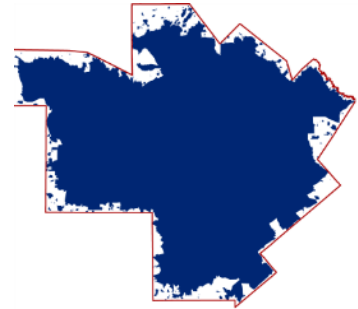

Road Density Map

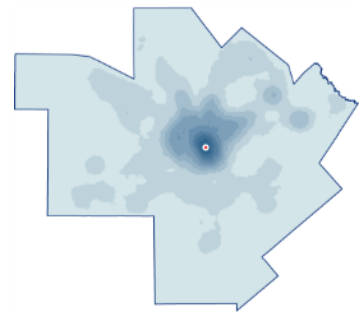

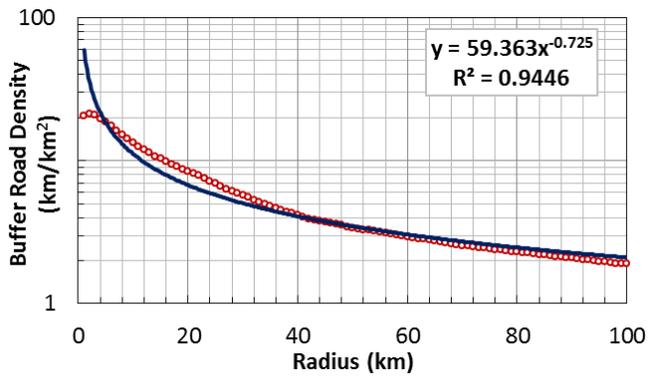

Road Density Fit

### Characteristics

| | |
|---|---:|
| Founded in | 1718 |
| Population | 2239307 |
| Pop Density (/km$^2$) | 138.1 |
| Area (km$^2$) | 16213.5 |
| Road Length (km) | 44137.5 |
| # of Intersections | 127773 |
| Area Threshold (m) | 875 |
| Line Threshold (m) | 806 |
| Point Threshold (m) | 347 |
| Density Index (km$^2$) | 59.363 |
| Decay Index (1/km) | 0.725 |

# San Diego, CA

Road Network

Road Polygon Area

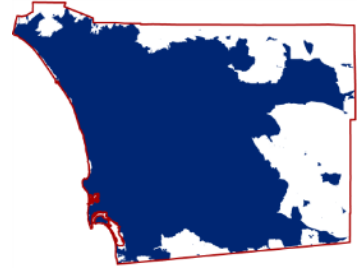

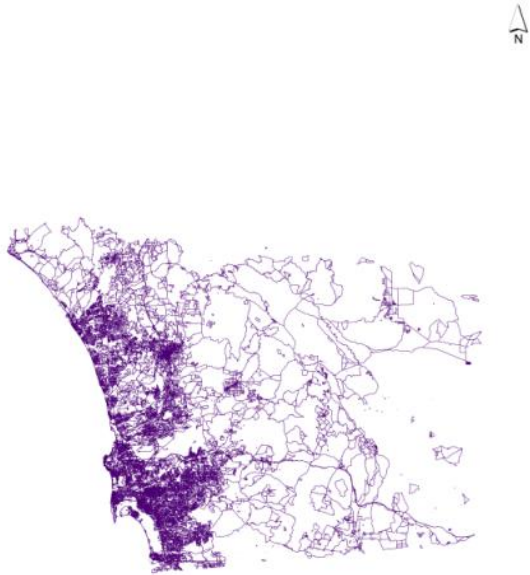

Road Density Map

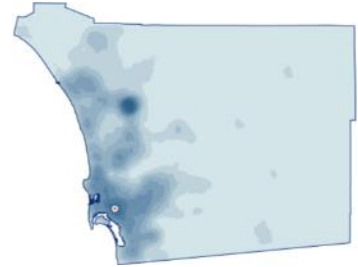

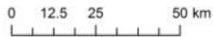

San Diego, CA

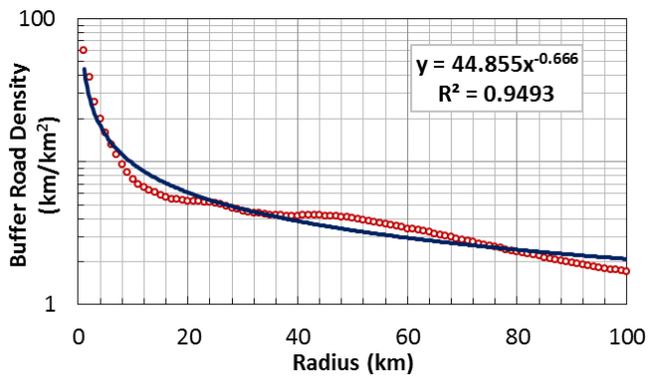

Road Density Fit

### Characteristics

| | |
|---|---:|
| Founded in | 1769 |
| Population | 3144425 |
| Pop Density (/km$^2$) | 410.1 |
| Area (km$^2$) | 7668 |
| Road Length (km) | 29499.1 |
| # of Intersections | 144194 |
| Area Threshold (m) | 1129 |
| Line Threshold (m) | 413 |
| Point Threshold (m) | 186 |
| Density Index (km$^2$) | 44.855 |
| Decay Index (1/km) | 0.666 |

# San Francisco, CA

Road Network

Road Polygon Area

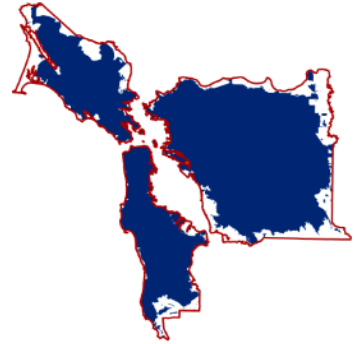

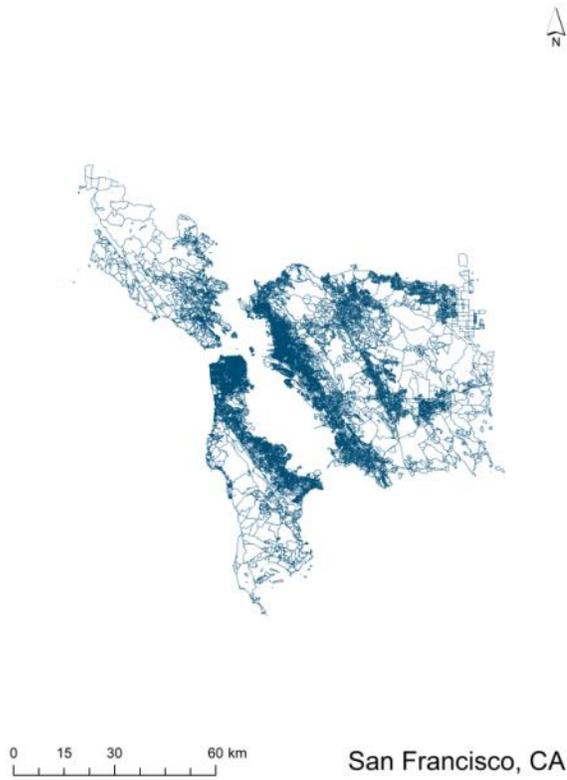

Road Density Map

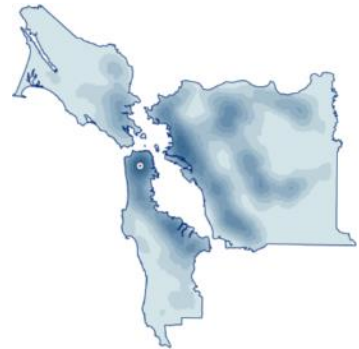

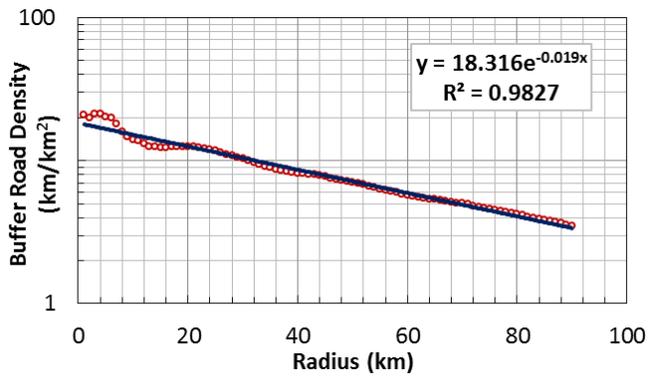

Road Density Fit

### Characteristics

| | |
|---|---:|
| Founded in | 1776 |
| Population | 4472992 |
| Pop Density (/km$^2$) | 835.7 |
| Area (km$^2$) | 5352.1 |
| Road Length (km) | 33483 |
| # of Intersections | 172400 |
| Area Threshold (m) | 640 |
| Line Threshold (m) | 272 |
| Point Threshold (m) | 155 |
| Density Index (km$^2$) | 18.316 |
| Decay Index (1/km) | 0.019 |

# San Jose, CA

Road Network

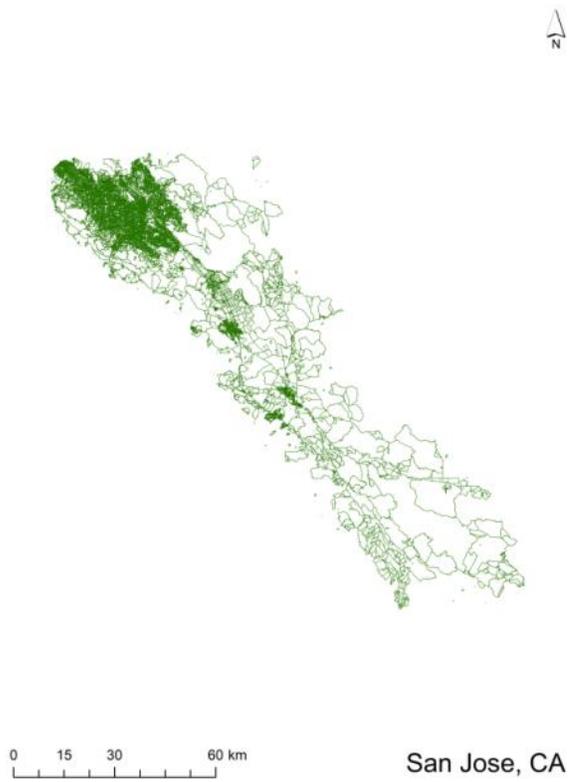

San Jose, CA

Road Polygon Area

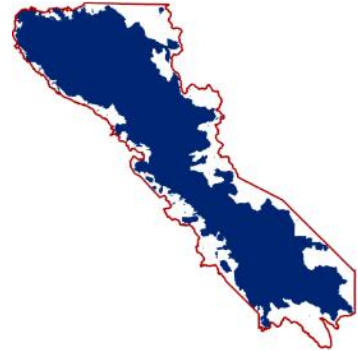

Road Density Map

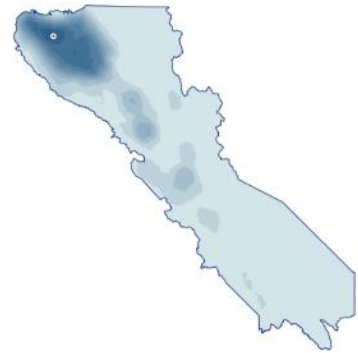

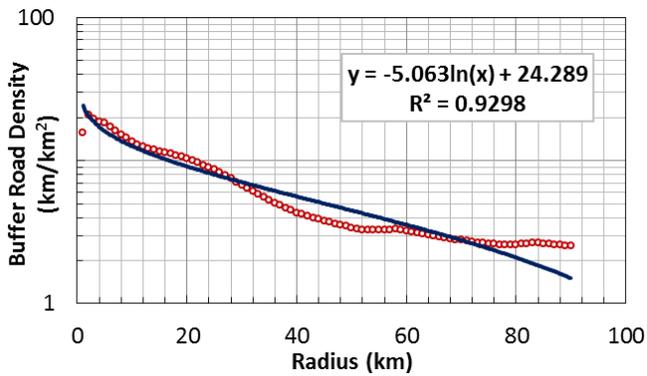

Road Density Fit

### Characteristics

| | |
|---|---:|
| Founded in | 1777 |
| Population | 1992872 |
| Pop Density (/km$^2$) | 405 |
| Area (km$^2$) | 4921.2 |
| Road Length (km) | 19824.6 |
| # of Intersections | 93610 |
| Area Threshold (m) | 773 |
| Line Threshold (m) | 478 |
| Point Threshold (m) | 195 |
| Density Index (km$^2$) | 24.289 |
| Decay Index (1/km) | 5.063 |

# St. Louis, MO

Road Network

Road Polygon Area

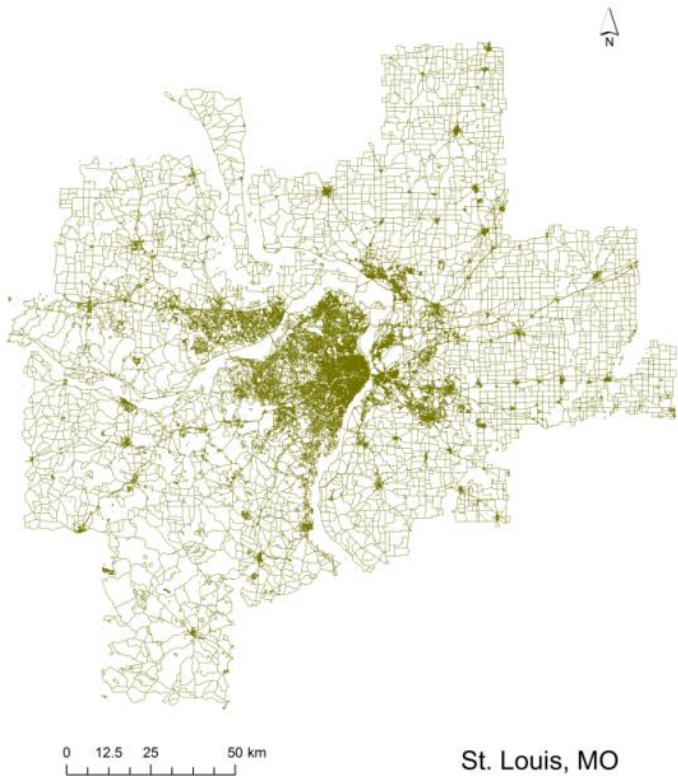

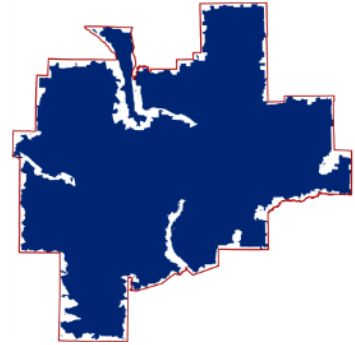

Road Density Map

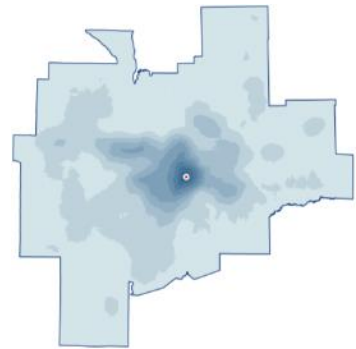

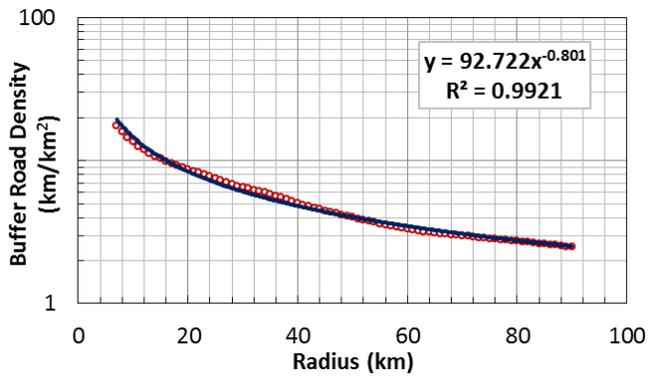

Road Density Fit

## Characteristics

| | |
|---|---:|
| Founded in | 1763 |
| Population | 2934412 |
| Pop Density (/km$^2$) | 145.4 |
| Area (km$^2$) | 20184.1 |
| Road Length (km) | 57670.8 |
| # of Intersections | 205269 |
| Area Threshold (m) | 880 |
| Line Threshold (m) | 753 |
| Point Threshold (m) | 287 |
| Density Index (km$^2$) | 92.722 |
| Decay Index (1/km) | 0.801 |

# **Tampa, FL**

Road Network

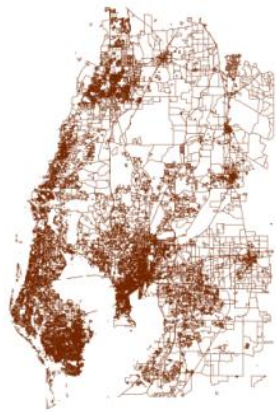

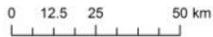

Road Polygon Area

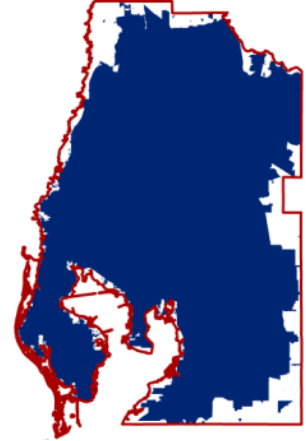

Road Density Map

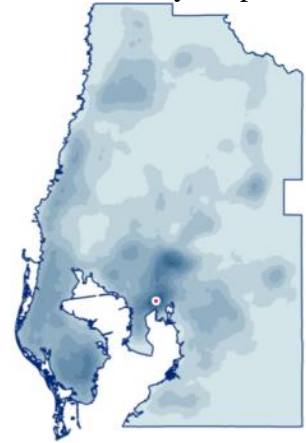

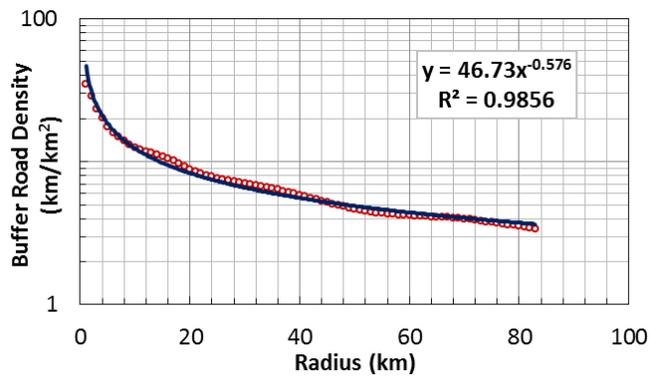

Road Density Fit

### **Characteristics**

| | |
|---|---:|
| Founded in | 1823 |
| Population | 2858974 |
| Pop Density (/km$^2$) | 496.6 |
| Area (km$^2$) | 5756.8 |
| Road Length (km) | 31421.2 |
| # of Intersections | 143714 |
| Area Threshold (m) | 756 |
| Line Threshold (m) | 315 |
| Point Threshold (m) | 180 |
| Density Index (km$^2$) | 46.73 |
| Decay Index (1/km) | 0.576 |

$y = 46.73x^{-0.576}$
$R^2 = 0.9856$

# **Washington, DC**

Road Network

Road Polygon Area

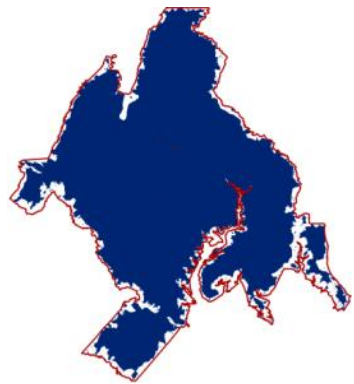

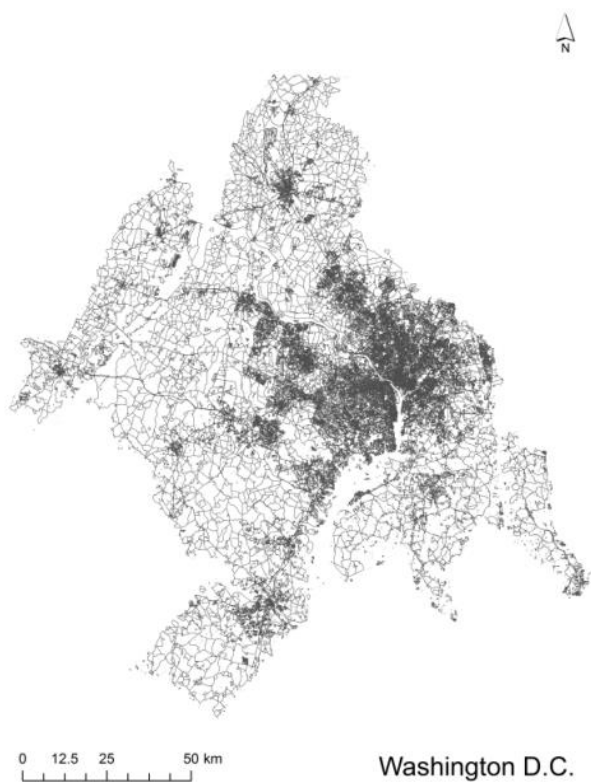

Road Density Map

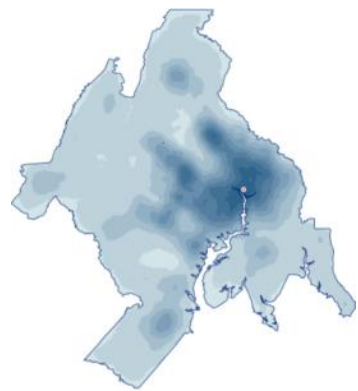

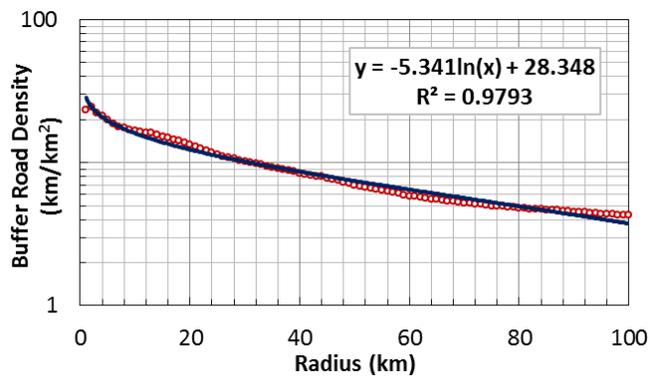

Road Density Fit

### **Characteristics**

| | |
|---|---:|
| Founded in | 1790 |
| Population | 5916033 |
| Pop Density (/km$^2$) | 464.5 |
| Area (km$^2$) | 12735 |
| Road Length (km) | 74190.6 |
| # of Intersections | 437470 |
| Area Threshold (m) | 467 |
| Line Threshold (m) | 361 |
| Point Threshold (m) | 162 |
| Density Index (km$^2$) | 28.348 |
| Decay Index (1/km) | 5.341 |